\documentclass[%
 pra,
 amsmath,amssymb,
 reprint,%
superscriptaddress]{revtex4-1}

\usepackage{graphicx}
\usepackage{dcolumn}
\usepackage{bm}
\usepackage{chemarrow}
\usepackage{color}
\usepackage{enumitem}

\draft 

\newcommand{\state}[4]{%
${#1}\, ^{#2}\mathrm{#3}_{#4}$}%
\newcommand{\statep}[4]{%
${#1}\, ^{#2}\mathrm{#3}_{#4}^{+}$}%

\newcommand{\Bstate}{%
\statep{B}{1}{\Sigma}{u} }
\newcommand{\HHstate}{%
\statep{HH}{1}{\Sigma}{g} }
\newcommand{\GKstate}{%
\statep{GK}{1}{\Sigma}{g} }
\newcommand{\FiveS}{%
\statep{5}{1}{\Sigma}{u} }
\newcommand{\SixS}{%
\statep{6}{1}{\Sigma}{u} }
\newcommand{\GrndS}{%
\statep{X}{1}{\Sigma}{g} }
\newcommand{\SigmaUS}{%
\statep{1}{2}{\Sigma}{u}(2p$\sigma_u$) } 
\newcommand{\SigmaGS}{%
\statep{1}{2}{\Sigma}{g}(1s$\sigma_g$) } 
\newcommand{\Cstate}{%
\state{C}{1}{\Pi}{u}(2p$\pi$) } 
\newcommand{\Dstate}{%
\state{D}{1}{\Pi}{u}(3d$\pi$) } 
\newcommand{\Dstatep}{%
\state{D'}{1}{\Pi}{u}(4p$\pi$) } 
\newcommand{\FiveP}{%
\state{5}{1}{\Pi}{u} }
\newcommand{\SixP}{%
\state{6}{1}{\Pi}{u} }
\newcommand{\Istate}{%
\state{I}{1}{\Pi}{g} } 
\newcommand{\Bprimestate}{%
\statep{B'}{1}{\Sigma}{u} } 
\newcommand{\Xplusstate}{%
\statep{X}{2}{\Sigma}{g}(1s$\sigma_g$) } 

\begin{document}

\preprint{AIP/123-QED}

\title[Non-Equilibrium Dynamics in Two-Color, Few-Photon Dissociative Excitation and Ionization of D$_2$] {Non-Equilibrium Dissociative Dynamics of D$_2$ in Two-Color, Few-Photon Excitation and Ionization}

\author{D.S. Slaughter}
\email{DSSlaughter@lbl.gov}
\affiliation{ 
Chemical Sciences Division, Lawrence Berkeley National Laboratory, 1 Cyclotron Rd, Berkeley, CA 94720, USA}

\author{F.P. Sturm}
\affiliation{ 
Chemical Sciences Division, Lawrence Berkeley National Laboratory, 1 Cyclotron Rd, Berkeley, CA 94720, USA}

\author{Roger Y. Bello}
\affiliation{ 
Chemical Sciences Division, Lawrence Berkeley National Laboratory, 1 Cyclotron Rd, Berkeley, CA 94720, USA}

\author{K.A. Larsen}
\affiliation{ 
Chemical Sciences Division, Lawrence Berkeley National Laboratory, 1 Cyclotron Rd, Berkeley, CA 94720, USA}
\affiliation{%
Graduate Group in Applied Science and Technology, University of California, Berkeley, CA 94720, USA}

\author{N. Shivaram}
\affiliation{ 
Chemical Sciences Division, Lawrence Berkeley National Laboratory, 1 Cyclotron Rd, Berkeley, CA 94720, USA}

\author{C.W.~McCurdy}
\affiliation{ 
Chemical Sciences Division, Lawrence Berkeley National Laboratory, 1 Cyclotron Rd, Berkeley, CA 94720, USA}
\affiliation{%
Department of Chemistry, University of California, Davis, CA 95616, USA}%

\author{R.R.~Lucchese}
\affiliation{ 
Chemical Sciences Division, Lawrence Berkeley National Laboratory, 1 Cyclotron Rd, Berkeley, CA 94720, USA}

\author{L. Martin}
\affiliation{ 
JILA and Department of Physics, University of Colorado and NIST, Boulder, CO 80309, USA}

\author{C.W. Hogle}
\affiliation{ 
JILA and Department of Physics, University of Colorado and NIST, Boulder, CO 80309, USA}

\author{M.M. Murnane}
\affiliation{ 
JILA and Department of Physics, University of Colorado and NIST, Boulder, CO 80309, USA}

\author{H.C. Kapteyn}
\affiliation{ 
JILA and Department of Physics, University of Colorado and NIST, Boulder, CO 80309, USA}

\author{P. Ranitovic}
\affiliation{ 
Chemical Sciences Division, Lawrence Berkeley National Laboratory, 1 Cyclotron Rd, Berkeley, CA 94720, USA}
\affiliation{ 
JILA and Department of Physics, University of Colorado and NIST, Boulder, CO 80309, USA}

\author{Th. Weber}
\email{TWeber@lbl.gov}
\affiliation{ 
Chemical Sciences Division, Lawrence Berkeley National Laboratory, 1 Cyclotron Rd, Berkeley, CA 94720, USA}

\date{\today}

\begin{abstract}
D$_2$ molecules, excited by linearly cross-polarized femtosecond extreme ultraviolet (XUV) and near-infrared (NIR) light pulses, reveal highly structured D$^+$ ion fragment momenta and angular distributions that originate from two different 4-step dissociative ionization pathways after four photon absorption (1 XUV + 3 NIR). We show that, even for very low dissociation kinetic energy release $\le$~240~meV, specific electronic excitation pathways can be identified and isolated in the final ion momentum distributions. With the aid of {\it ab initio} electronic structure and time-dependent Schr\"odinger equation calculations, angular momentum, energy, and parity conservation are used to identify the excited neutral molecular states and molecular orientations relative to the polarization vectors in these different photoexcitation and dissociation sequences of the neutral D$_2$ molecule and its D$_2^+$ cation. In one sequential photodissociation pathway, molecules aligned along either of the two light polarization vectors are excluded, while another pathway selects molecules aligned parallel to the light propagation direction. The evolution of the nuclear wave packet on the intermediate \Bstate electronic state of the neutral D$_2$ molecule is also probed in real time. 
\end{abstract}

\pacs{78.47.J−, 25.60.Gc, 42.50.Tx, 33.80.Gj, 33.80.Rv, 34.80.Gs, 34.80.Ht, 82.50.Pt, 82.53.−k, 82.53.Eb}
\keywords{ultrafast, photoexcitation, photoionization, photodissociation, particle momentum imaging }
\maketitle

\section{\label{intro}Introduction}

The hydrogen molecule H$_2$ is the simplest neutral molecule consisting of only two protons and two electrons. Due to its relative simplicity, H$_2$ serves as a model candidate for studying photo-excitation and ionization dynamics. In particular, one challenge for modern multi-dimensional spectroscopy is to understand how the dissociation dynamics far away from equilibrium evolves in time~\cite{hemminger_challenges_2015} and how the relative  populations of specific excited states can be controlled. In recent years, sporadic progress for small molecular systems has been made~\cite{sandhu_observing_2008,zhou_probing_2012,champenois_involvement_2016,lepine_molecular_2013,cao_identification_2015,sturm_mapping_2017}. More insight is pressing, since understanding the excited state behavior of electrons and nuclei, as well as their interplay, in simple fundamental systems is a crucial step towards understanding and controlling photon-driven processes in more complex molecules. In this work, the step-wise dissociation of the simplest homonuclear diatomic molecule H$_2$ is investigated through a detailed analysis of the fragment ion momentum distributions. 

Optical transitions to excited electronic states of the hydrogen molecule can be found within the vacuum ultraviolet spectrum, and the nuclear wavepacket dynamics occur on ultrafast timescales (tens of fs), creating challenging experimental requirements for detailed measurements of this fundamental molecule using standard spectroscopic tools of femtochemistry~\cite{zewail_femtochemistry_2000,sarkisov_femtochemistry_2001}. For example, in a 2-photon pump-probe experiment, photon energies higher than the 7$^{th}$ harmonic of a conventional Ti:sapphire $\thicksim$ 800~nm NIR laser are required, in order to access the neutral excited electronic states of H$_2$. 
Furthermore, in order to probe dynamics on excited electronic states of H$_2$, ultrashort vacuum ultraviolet pulse durations on the order of tens of femtoseconds are necessary. These photon energies and pulse durations are not achievable using standard nonlinear optical techniques such as sum frequency generation in gases or nonlinear crystals. The more recent field of attosecond physics, on the other hand, has primarily focused on the use of 
broadband XUV pulses, either in the form of attosecond pulse trains or isolated attosecond pulses~\cite{calegari_advances_2016,drescher_communication_2016,johnsson_attosecond_2007} produced by high harmonic generation (HHG) of laser light. Due to the challenges in producing spectrally-isolated few femtosecond pulses with sufficiently high fluence, the measurement of neutral excited state dynamics of the H$_2$ molecule in a pump-probe scheme has hence remained a challenge for both attophysics and femtochemistry communities.

The hydrogen molecule is an ideal test bed for developing ultrafast spectroscopy methods drawn from both femtochemistry and attophysics. Only recently it was shown that below-ionization-threshold harmonics (i.e. the 7$^{th}$ and the 9$^{th}$ harmonics) can be used to coherently control the dynamics of neutrally excited H$_2$ molecules on an attosecond timescale~\cite{ranitovic_attosecond_2014}. Moreover, H$_2$ serves as a fundamental and prototypical target for theoretical investigations and models. Understanding neutral excited state dynamics, even in the simple excited H$_2$ molecule, still represents a theoretical and experimental challenge due to the complex structure of high-lying excited electronic states. Rich information can be extracted from fragment angular distributions of the dissociation, which are {\it inter alia} very sensitive to selection rules, and allow for specific transitions to be identified in order to test theoretical treatments of two-color few-photon investigations \cite{Larsen}. In this report, we show that coherent XUV and NIR pulses can be used in a two-color multi-photon absorption scheme to initiate sequential dissociation dynamics. Cross-polarized XUV and NIR pulses are used to first excite the neutral H$_2$ molecule in different photoabsorption sequences, followed by ionization near threshold (with excess energies $<$ 450~meV). We applied 3D ion momentum imaging of the fragment proton to unravel the complex dissociative dynamics, induced by the cross-polarized two-color fields.

The electronic structures of H$_2$ and D$_2$ are identical. We hence used molecular deuterium gas as a target instead of molecular hydrogen, because our 3D ion momentum imaging setup allows us to distinguish D$^+$ fragments easily from H$^+$, which can be produced from vacuum contaminants, by their mass separation in time of flight. However, the vibrational levels of hydrogen and deuterium molecules differ by a few 100~meV, which can result in different ion spectra. The lowest vibrational level $\nu$ = 0 of the ground state is slightly higher for hydrogen molecules (0.275~eV) than for deuterium molecules (0.197~eV)~\cite{fantz_franckcondon_2006}. As the Potential Energy Curves (PECs) shown here are set to zero for the $\nu$ = 0 level of the \GrndS ground state of neutral hydrogen molecules, the equivalent curves for deuterium molecules are shifted up by 0.079~eV. The single ionization threshold for H$_2$ and D$_2$ molecules is 15.4~eV, while the dissociative threshold is $\sim$18.1~eV, meaning that for an electronic excitation with the 9$^{th}$ order harmonic light pulse, three additional $\sim$800~nm NIR photons are required to dissociate the molecular cation.

\section{\label{exp}Experimental methods}

The experimental setup is similar to that employed previously~\cite{sturm_mapping_2017}, with the addition of a low-pass filter gas cell enabling the isolation of XUV photon energies below 14~eV. In the present experiments, ultrashort near-infrared (NIR, 805~nm, 1.55~eV) laser pulses are produced by a Ti:sapphire oscillator (KM Labs), which are stretched and amplified, first by a Ti:sapphire regenerative amplifier (Positive Light Legend) and then a 6-pass Ti:sapphire amplifier at a 50~Hz repetition rate (see~\cite{sturm_time_2016}). The amplified pulses are compressed in a reflective grating compressor, which is housed in a vacuum chamber filled with 100~Torr of helium gas. The helium gas allows for effective cooling of the optics illuminated by the high intensity laser beam. 
The resultant compressed pulses, having 25~mJ of energy and $\sim$45~fs duration, are transported through a 400~$\mu$m-thick anti-reflective coated UV fused silica window from the helium filled chamber into ultrahigh vacuum. The linear polarization direction of the amplified NIR pulses is controlled by a 400~$\mu$m-thick crystal quartz half-wave plate. The NIR beam is focused by a 6~m focal length curved mirror (f/200) into a 10~cm long gas cell containing argon gas, at a pressure of 8 Torr, to create vacuum (VUV) and extreme ultraviolet (XUV) photons by High-order Harmonic Generation (HHG) of the fundamental 805~nm NIR wavelength of the driving laser. The gas cell is confined by stainless steel shims, with apertures drilled by the very same high intensity NIR beam. To minimize losses or temporal dispersion, the half-wave plate and the thin fused silica window are the only two transmissive optics following compression. 

After the HHG gas cell, the co-propagating XUV and NIR pulses enter a 1~m long argon gas cell at a pressure of approximately 3~Torr, which suppresses high order harmonics above the 9$^{th}$ order. A flat silicon mirror at Brewster's angle for the p-polarized NIR beam suppresses the fundamental wavelength relative to the high-order harmonics, which are reflected with an efficiency of $>$60~\%. A second silicon mirror at Brewster's angle further suppresses any residual p-polarized NIR component of the incident light beam, while transporting the XUV beam with the same high efficiency as the first mirror. Fine adjustment of the half-wave plate enables the precise control of s-polarized NIR light to co-propagate with the XUV pulse. 
By adjusting the pressure in the HHG gas cell, as well as the pressure in the argon gas filter, an efficient population of the neutral states, and simultaneous suppression of the direct ionization of the D$_2$ target molecules in the experimental vacuum chamber downstream, is achieved. 

The photon beam propagates along the lab-frame x-axis and through the main experimental vacuum chamber and reflects from a near-normal incidence ($< 3^{\circ}$) back-focusing mirror (f = 15 cm). The focus is aligned to a COLd Target Recoil Ion Momentum Spectroscopy (COLTRIMS) apparatus, consisting of a narrow, collimated two-stage supersonic gas jet of molecular deuterium, propagating vertically (y-axis), and a 3D momentum imaging ion spectrometer with the Time-Of-Flight (TOF) z-axis orthogonal to both the photon beam (x-axis) and the jet propagation direction (y-axis)~\cite{dorner_cold_2000,ullrich_recoil-ion_2003}. The deuterium molecules, emerging from an adiabatic expansion, have temperatures $<$~80~K, therefore their vibrationally excited states have negligible population. The temporal width of the 9$^{th}$ harmonic is approximately 15~fs, significantly shorter than the 45~fs NIR pulse duration. The spectral width of the NIR and XUV pulses are $\sim$80~meV and $\sim$300~meV, respectively. The intensity of the NIR beam is estimated to be on the order of approximately 3 $\cdot$ 10$^{11}$ W/cm$^2$, based on an estimated suppression factor of $\geq$ 2 $\cdot$ 10$^{-5}$, obtained from the two silicon mirror reflections at Brewster's angle. With these intensities we expect resonant few-photon excitation by the NIR and negligible 2-photon resonant excitation by the XUV pulse to take place. We note that there are no excited electronic states below 11~eV in D$_2$. Therefore, electronic excitation must involve the XUV light pulse, specifically the 9$^{th}$ harmonic.

The ions created in the interaction region of the 3D ion momentum imaging spectrometer, spanned by the overlap of our gas jet of around 2~mm in diameter and the light focus ($\thicksim$ 0.006 $\cdot$ 0.012 $\cdot$ 0.22~mm), are guided along an extraction region of 7.1~cm length by a homogeneous electric field of 4~V/cm towards a time-sensitive 120~mm multi-channel plate detector, equipped with a dual-layer delay-line anode for position readout~\cite{roentdek_mcp_nodate}. The ion energy resolution, derived from a calibration measurement on vibrational states to the L1 dissociation limit of oxygen molecules after single ionization at 23.25~eV~\cite{ranitovic_attosecond_2018}, amounts to $\Delta$E$_{D^+}$ $\le$ $\pm$15~meV. The simulated azimuthal ion angular resolution is $\Delta\phi_{D^+}$ $\le$ $\pm$ 5$^{\circ}$. The experiment was carried out on a shot-by-shot basis and recorded in list-mode file format, allowing for an intricate offline analysis, including retrieving, sorting, and extracting the relevant data and momentum calibration. The TOF of the ions (along the z-axis, i.e. the XUV polarization direction) and the impact positions on the detector in two dimensions (along the x-axis, i.e. the light propagation direction, and along the y-axis, i.e. the gas jet propagation direction and NIR polarization direction) are used to retrieve the final state ion momenta in three dimensions. Only the charged D$^+$ ion is measured, while the neutral D atom in the dissociation process is undetected. However, since the light pulses are ultrashort and the dissociation dynamics in this energy regime are on the order of tens of femtoseconds, no rotation of the hydrogen molecule upon excitation and ionization before the fragmentation is expected, i.e. the axial recoil approximation holds, and the back-to-back emission of the nuclear fragments D$^+$ and D in the dissociation resembles the orientation of the molecular axis at the instant of photoabsorption. The laboratory coordinate frame is defined by the XUV and NIR polarization as well as the light propagation direction, which are approximately parallel to the spectrometer z (TOF and XUV polarization direction), y (gas jet and NIR polarization direction), and x (XUV and NIR light propagation) axes, respectively.

\section{\label{theory}Theoretical methods}

The PECs and dipole couplings between 
H$_2$ neutral states, as a function of the internuclear 
distance, were computed using the multi-reference 
configuration interaction (MRCI) capability of MOLPRO \cite{werner_molpro_2012, werner_molpro_2015}, with single and double excitations from an active space. Due to the distinct characters of the states involved, different active spaces were used for different gerade-ungerade groups of states \cite{SPIELFIEDEL2003162}. The following convention was adopted for each of these groups:
$^1\Sigma_g^+$-$^1\Sigma_u^+$: $11\sigma_g$, $8\sigma_u$, $3\pi_g$, and $3\pi_u$; $^1\Sigma_g^+$-$^1\Pi_u$: $11\sigma_g$, $5\sigma_u$, $3\pi_g$, and $5\pi_u$; $^1\Sigma_u^+$-$^1\Pi_g$: $8\sigma_g$, $8\sigma_u$, $4\pi_g$, and $4\pi_u$; $^1\Pi_g$-$^1\Pi_u$: $8\sigma_g$, $5\sigma_u$, $4\pi_g$, and $5\pi_u$.
The one-electron basis 
set was aug-cc-pVTZ \cite{Dunning_1989, Dunning_1992}, augmented by eight $s$, eight $p$, eight $d$, and four $f$ diffuse functions \cite{Kaufmann} centered at the bond midpoint. 

The state population as a function of delay between separate XUV and NIR pulsed fields were calculated by solving the Time-Dependent Sch\"odinger-Equation (TDSE). The time-dependent wave function was expanded in a basis of Born-Oppenheimer vibrational states \cite{Bello_2021}, which were found by diagonalizing the nuclear Hamiltonian using the PECs discussed above. 
In order to simulate the different paths probed by the experiment two separated calculations were performed. In the first one (shown in Fig.~\ref{theoryfig} panels (a) and (c)), only states having $^1\Sigma_g^+$, $^1\Sigma_u^+$, and $^1\Pi_g$ symmetries were included, while in the second one (shown in Fig.~\ref{theoryfig} panels (b) and (d)) states having $^1\Sigma_g^+$, $^1\Pi_u$, and $^1\Pi_g$ symmetries were included. All the calculations were performed considering molecules oriented 45$^\circ$ with respect to the XUV polarization vector.  

\section{\label{results}Results}

\begin{figure}
\includegraphics[width=1.0 \columnwidth]{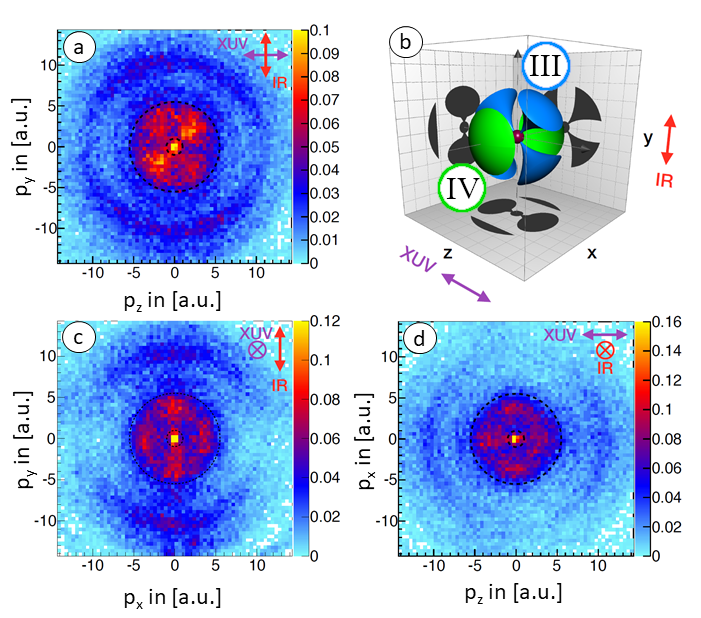}
\caption{\label{momenta}
D$^+$ ion momentum distributions in all spatial directions for angular cuts $|\cos(\theta_i)|\leq 0.84$ (see text). The inner and outer dashed circle indicate ion kinetic energies of 10~meV and 130~meV, respectively. (a) D$^+$ momentum in the NIR/XUV polarization (y-z) plane. (b) sketch of feature (III) in blue and feature (IV) in green (see text). (c) ion momentum distribution in the NIR-polarization/light-propagation-direction (y-x plane). (d) ion momentum distribution in the light-propagation-direction/XUV-polarization (x-z plane).}
\end{figure}

A slice of the detected D$^+$ ion momentum distribution of the molecular dissociation is shown in Fig.~\ref{momenta}(a). The slice is defined by an angular cut of $|cos(\theta_i)|\leq 0.84$, with $\theta_i$ being the angle between the momentum vector and the plane i under observation, which in this case is the y-z plane. This is the plane parallel to the TOF (z-axis) and the gas jet (y-axis), which is orthogonal to the laser propagation direction (x-axis), within an acceptance angle of $\theta_i \leq \pm 33^\circ$. 
In this representation the XUV polarization vector is orientated horizontally along the z-axis, and the NIR polarization vector is orientated vertically along the y-axis, as indicated in the small inset in Fig.~\ref{momenta}(a). The momentum distribution in this NIR/XUV polarization y-z plane reveals three contributions: (I) a ring with maximum intensity along the NIR direction and a radius of around 10 atomic units (a.u.) in D$^+$ ion momentum, which corresponds to 0.44~eV in kinetic energy; and (II) a faint second distribution with a radius of around 6~a.u. corresponding to 0.18~eV. 
(III) an intense 4-lobed pattern, resembling a d-wave spherical harmonic or a cloverleaf, within a maximum radius of 5~a.u., as indicated by the dashed black circle in Fig.~\ref{momenta}(a), corresponding to D$^+$ kinetic energies $\leq$ 120~meV. 
The bright spot within the small dashed black circle of 1~a.u. in momentum, corresponding to 10~meV in kinetic energy, is a result of the angular slice combined with the finite bin size of the spectrum and is not statistically relevant. We will neglect this artifact in the following analysis and discussion. 
To guide the eye, we provide a schematic depiction of the relevant ion momentum distributions in Fig.~\ref{momenta}(b). The 4-lobed D$^+$ ion momentum distribution in three dimensions of feature (III) is sketched in blue, while the dipole shaped emission pattern, which we call feature (IV), is sketched in green.

Features (I) and (II) are known to stem from direct XUV + 2 NIR ionization processes. H$^+$ and D$^+$ ions with energies around 0.18 and 0.44~eV have been observed previously in two color measurements on the photodissociaton of H$_2$ and D$_2$ molecules~\cite{sturm_mapping_2017,kelkensberg_molecular_2009,furukawa_nonlinear_2010,cao_identification_2015}. They are generally attributed to populating the H$_2^+$/D$_2^+$ cation ground state \SigmaGS by one XUV photon (11$^{th}$ harmonic = 17.05~eV) or a combination of an XUV photon and NIR pulses (i.e. 9$^{th}$ harmonic + 2 NIR = 13.95 + 3.1~eV) followed by a parallel transition to the dissociative \SigmaUS cation state via the absorption of one additional NIR photon~\cite{furukawa_nonlinear_2010}. This transition represents the bond softening dissociation process after single photoionization, discussed in e.g. references~\cite{sturm_mapping_2017,kelkensberg_molecular_2009,cao_identification_2015,posthumus_dynamics_2004}. In this picture the electronic states of the molecule are dressed by the NIR field, which lowers the dissociation potential energy barrier by an amount depending on the available NIR intensity. This can result in a resonant coupling of the bound \SigmaGS D$_2^+$ state to the dissociative \SigmaUS D$_2^+$ cation state by a single NIR photon, which yields D$^+$ energies of 0.4 to 0.7 eV, as reported previously in Ref.~\cite{sturm_mapping_2017}. The 2-step photon absorption sequence is given by:\\

(I and II):
\begin{align*}
\textrm{\GrndS} \xrightarrow{XUV + 2{\cdot} NIR}~&\textrm{\SigmaGS} + \textrm{e}^{-}\\
\xrightarrow{NIR}~&\textrm{\SigmaUS} + \textrm{e}^{-}.
\end{align*}

The two complementary views of the three dimensional D$^+$ ion momentum distribution, in the y-x and x-z planes, are presented in Fig.~\ref{momenta}(c) and (d), respectively. As in Fig.~\ref{momenta}(a), these data are extracted by applying momentum slices in specific planes i (i $\in$ [x-z, y-x]) with an acceptance angle of $\theta_i \le \pm33^\circ$. Fig.~\ref{momenta}(c) shows the D$^+$ ion momentum yield in the y-x plane, including the NIR polarization direction (y-axis) and the photon propagation direction (x-axis). Fig.~\ref{momenta}(d) shows the D$^+$ ion momentum yield in the x-z plane including the photon propagation direction (x-axis) and the XUV polarization (also TOF direction, z-axis). In both Figs.~\ref{momenta}(c) and (d), we see an additional contribution, feature (IV), of D$^+$ fragment ions being emitted forward and backward along the photon propagation direction (x-axis) with a fixed momentum of around 3.5~a.u. within narrow cones, which are illustrated schematically as green cones in Fig.~\ref{momenta}(b). 
Note the momentum slices in Fig.~\ref{momenta}(c) and (d) show profiles of the cloverleaf structure of Fig.~\ref{momenta}(a), and, hence, the structures parallel to the XUV and NIR polarizations are projected perspectives of feature (III). Keeping in mind the finite acceptance angles of the applied momentum slices ($|\cos(\theta_i)| \le 0.84$), these peaks have minima in the polarization directions, as seen most clearly in Fig.~\ref{momenta}(a).

\section{\label{discussion}Discussion}

\begin{figure}
\begin{tabular}{ll}
\includegraphics[trim={0cm 11.75cm 15cm 0cm},clip,width=0.7 \columnwidth]{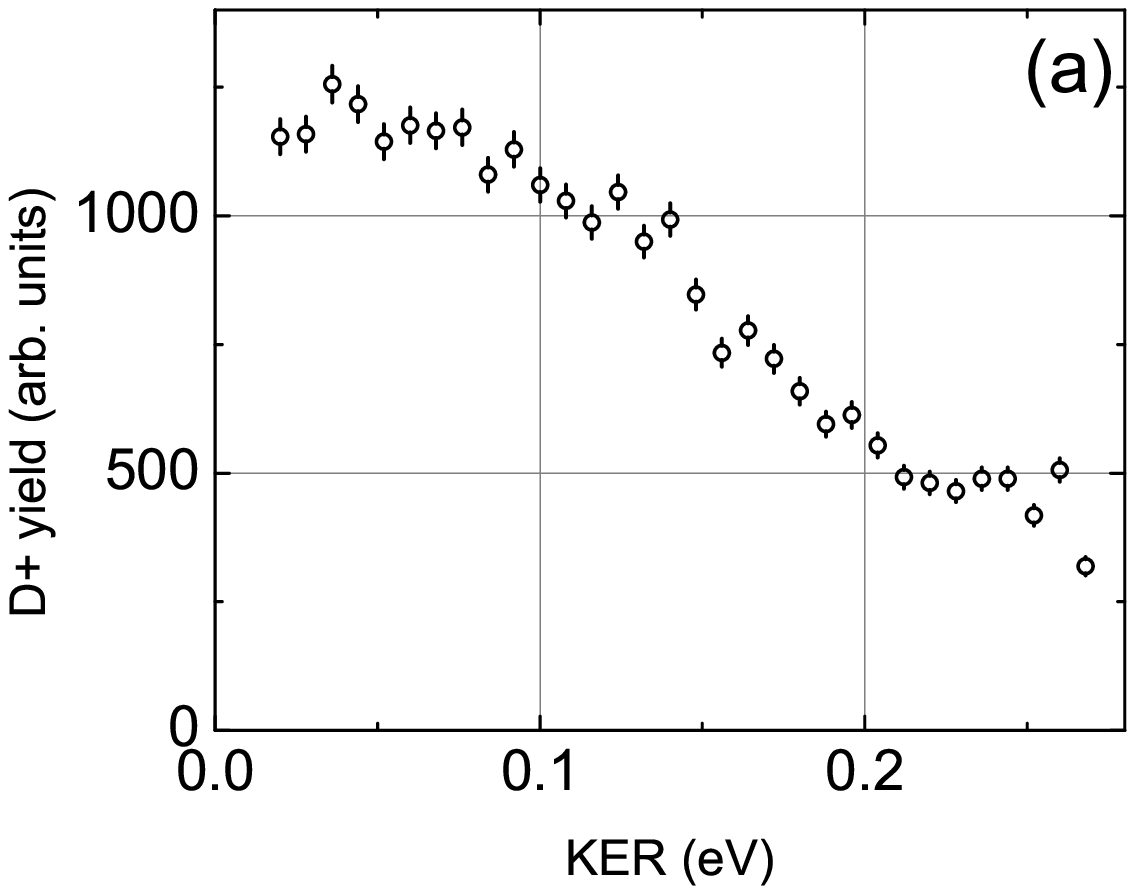}\\
\includegraphics[trim={0cm 10cm 15cm 1.2cm},clip,width=0.7 \columnwidth]{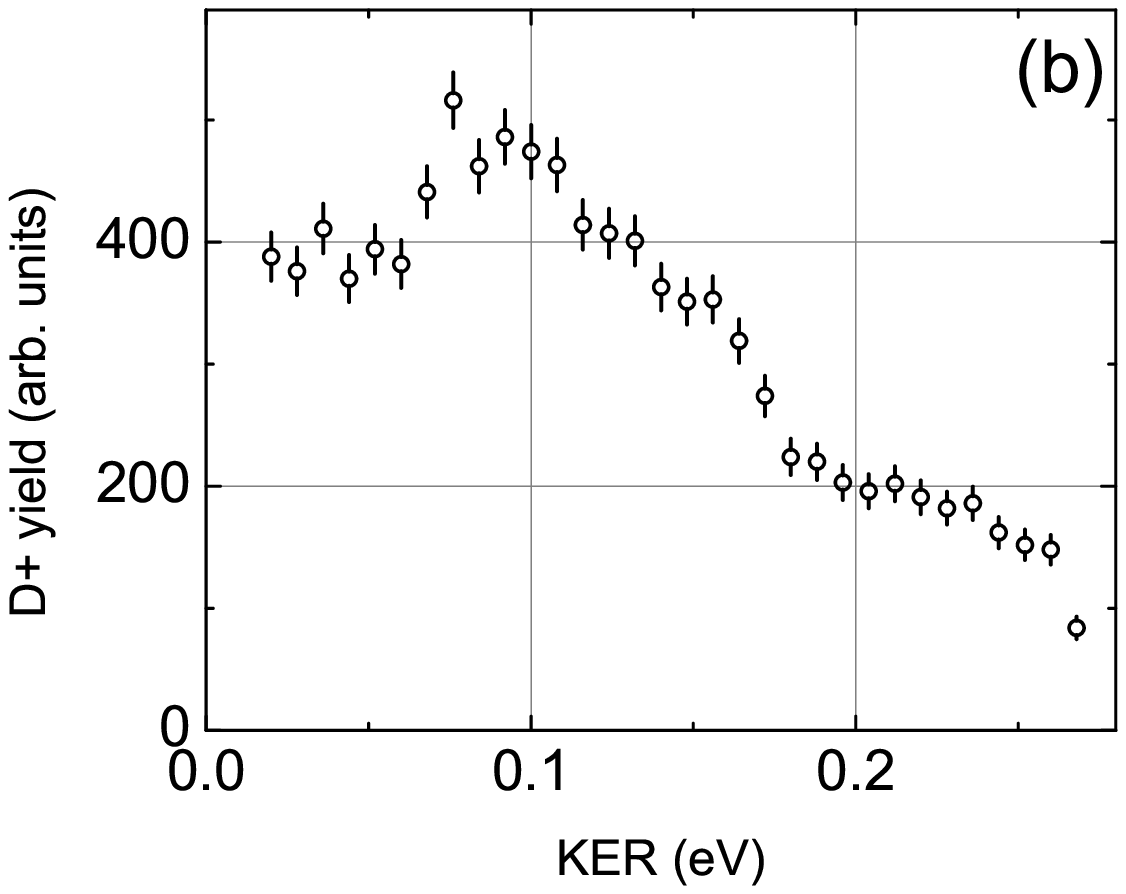}\\
\end{tabular}
\caption{\label{KERs}
(a) KER distribution of feature (III). (b) KER distribution of feature (IV). Only D$^+$ ions having momenta between 1~a.u. and 6~a.u. and emission angles within $\theta_i \le \pm33^\circ$ of the relevant plane i are included in each plot. In (b) an additional condition is applied, accepting only D$^+$ emission angles within a cone of $\pm$40$^\circ$ around the laser light propagation direction (x-axis). 
}
\end{figure}

\begin{figure}
\includegraphics[width=1.0 \columnwidth]{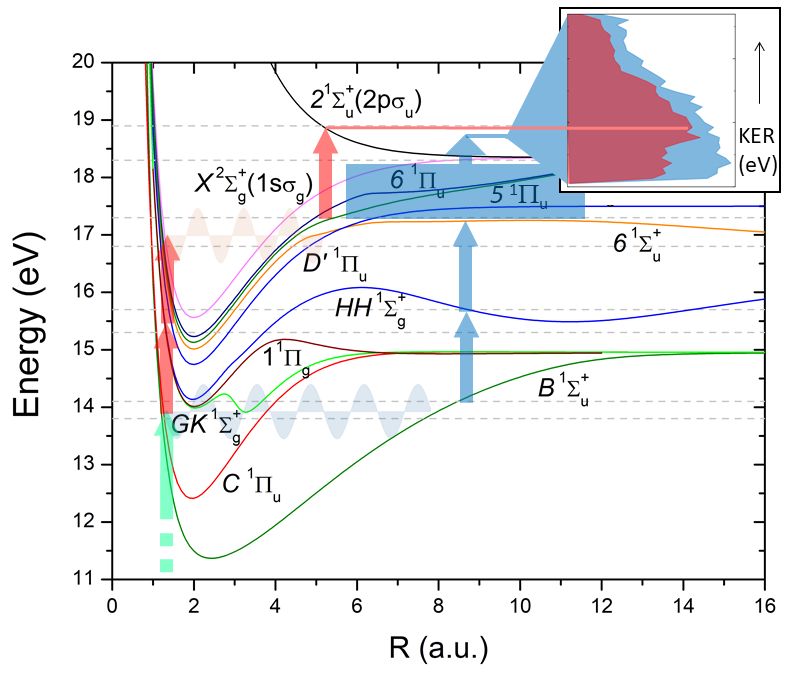}
\caption{\label{scheme}
Potential energy curves of molecular hydrogen and schematic representation of the dissociation and ionization steps. Neutral H$_2$ molecules are excited by a 13.95~eV XUV photon (green arrow) to the \Bstate and \Cstate states. On the \Bstate, a nuclear wavepacket is spawned, then two resonant 1.55~eV NIR photoexcitations (blue arrows) populate the \HHstate and highly excited \SixS states at larger internuclear distance R. Subsequent dissociative photoionization by a third 1.55~eV NIR photon over a wide range of internuclear distances (broad blue arrow) occurs on the \SigmaUS state. On the \Cstate state, two resonant  1.55~eV NIR photoexcitations (red arrows) populate the \HHstate/\Istate/\GKstate and the highly excited \FiveP and \SixP states. A subsequent dissociative photoionization by a third 1.55~eV NIR photon (red arrow) on the \SigmaUS state or a degenerate continuum vibrational level of the \SigmaGS cation state) occurs near R~$\approx$~5~a.u.. The \FiveP and \SixP Rydberg neutral states can only be accurately calculated to R~$\approx$~8~a.u. max. Dashed gray lines denote the available transition energies and bandwidth (full width at half maximum) given by the XUV (300~meV) and NIR (80~meV) pulses. The inset shows the zoomed KER distributions of features (III) in blue and (IV) in red, which are assigned to the \Bstate state and \Cstate pathways, respectively.  
}
\end{figure}

We will start the discussion with the ionization process producing feature (III) in Fig.~\ref{momenta}(b), by first looking at the energy balance of the fragmentation process. The dissociative single ionization threshold for H$_2$ molecules is $E_{diss}(H_2^+)=18.078$~eV~\cite{cattaneo_attosecond_2018} ,and for D$_2$ it is $E_{diss}(D_2^+)=18.158$~eV~\cite{balakrishnan_dissociation_1994}. The dissociation limit leading to the neutral dissociation H(1s) + H(2$l$) with $l=\in[s, p]$ is $E_{diss}(H_2)=14.67$~eV. Both limits are above the XUV photon energy used in the present experiment, i.e. a single XUV photoabsorption alone cannot trigger a fragmentation process by itself. However, the absorption of a few NIR photons following XUV photoabsorption enables the population of the repulsive molecular D$_2^+$ cation \SigmaUS state on which the molecule dissociates with a limit of $E_{diss}(D_2^+)$, producing the deuterium fragment ion D$^+$ with very low kinetic energy that we detect and a neutral fragment D. 
In principle, at this internuclear distance we can also populate a dissociative vibrational level of the \SigmaGS state, which is degenerate with the continuum of the \SigmaUS state. A D$^+$ ion with a maximum kinetic energy of 120~meV corresponds to a kinetic energy release (KER) of the dissociation of only 240~meV (see Fig.~\ref{KERs}(a)). The small KER of $\le$~240~meV is consistent with photoionization to the dissociative \SigmaUS state, a dissociative vibrational level of \SigmaGS, for the D$_2^+$ internuclear distance of $\ge$ 9.0 a.u.. This suggests the following 4-step dissociative ionization process (see Fig.~\ref{scheme}, green and blue arrows): \\

(III): 
\begin{align*}
\textrm{\GrndS} &\xrightarrow[res.]{XUV} ~\textrm{\Bstate}\\ &\xrightarrow[res.]{NIR} \textrm{\HHstate}\\
&\xrightarrow[res.]{NIR} \textrm{\SixS}\\ 
&\xrightarrow[cont.]{~NIR~} ~\textrm{\SigmaUS} + \textrm{e}^-.
\end{align*}

The 9$^{th}$ harmonic XUV possesses a central photon energy of 13.95~eV and can excite the \Bstate or \Cstate state, and also a few vibrational states of \Bprimestate and \Dstate, of the neutral D$_2$ molecule. Fig.~\ref{scheme} illustrates a nuclear vibrational wave-packet (NWP) launched in the \Bstate or \Cstate states (see Fig.~\ref{scheme}, green arrow). The \Bstate state can achieve large internuclear distances, enabling the observed low kinetic energies of the detected ions (Fig.~\ref{KERs}(a)), following a vertical transition from the ground electronic \GrndS state of the D$_2$ molecule. Excitation to the \Cstate state will be discussed later for pathway (IV). In the second and third step, the simultaneous absorption of two NIR photons can excite the NWP on the \Bstate state resonantly at large internuclear distances of 8.6~a.u. (4.55~\AA), near the outer turning point region of the \Bstate state, which consecutively populates the neutral \HHstate and without delay the \SixS state (see Fig.~\ref{scheme}, blue arrow).

\begin{figure}[h]
\includegraphics[trim={1.1cm 0.5cm 0cm 0cm},clip,width=1.0\columnwidth]{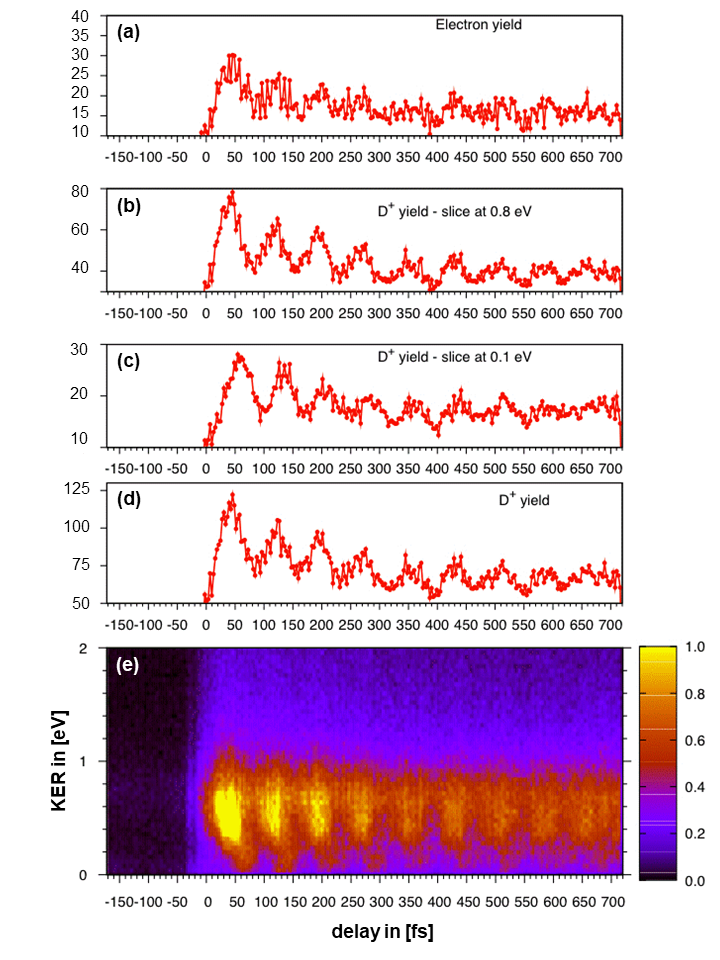}
\caption{\label{timeresolved}
Evolution of the NWP on excited states of D$_2$ molecule, measured by time-resolved electron yield (a), D$^+$ ion yield for a KER of 0.8~eV (b), D$^+$ ion yield for a KER of 0.1~eV (c), and D$^+$ ion yield for all energies. (e) KER of the D$^+$ + D fragmentation as a function of pump-probe delay.}
\end{figure}

This proposed excitation scheme is only possible if the NWP reaches the outer turning point within the NIR pulse duration, so that the NIR field is still present to further excite the molecule at this larger internuclear distance. As mentioned above, the temporal width of the XUV pulse is on the order of 15~fs full width at half maximum (FWHM), while the NIR pulse, which is simultaneously co-propagating with the XUV pulse, has a broader duration of 45~fs FWHM. This means that the maximum time for this sequential excitation to happen must be less than $\sim$ 60~fs. We conducted a separate experiment to measure the evolution of the NWP on the \Bstate state in real time, in order to understand how efficiently this sequential exaction step takes place in the temporal domain. This was realized with a 10~kHz, 25~Watt, 840~nm NIR laser system, which was driving high-order harmonic generation in a xenon filled gas cell. The 9$^{th}$ harmonic with an energy of 13.28~eV was accompanied by a weak NIR laser field (2$\cdot$10$^{11}$~W/cm$^2$) and followed by a stronger (5$\cdot$10$^{12}$~W/cm$^2$) time delayed NIR probe pulse. Note the lower photon energies of the 9$^{th}$ harmonic XUV and the NIR pulses in this time-resolved experiment. The XUV and NIR pulses were linearly polarized and parallel to each other. The pulse width of the XUV attosecond pulse train was between 5 and 10~fs, while the NIR pulse was around 30~fs in duration. The XUV pump pulses were focused by a pair of multi-layer mirrors, with an effective focal length of 1~m into a 3D electron and ion momentum imaging spectrometer, and the NIR probe pulses were focused by a spherical mirror of the same focal length. 3D momentum of the electron and D$^+$ recoil-ion were recorded in coincidence \cite{hogle_attosecond_2015}. The results are presented in Fig.~\ref{timeresolved}. The spectrum in Fig.~\ref{timeresolved}(e) depicts the measured KER of the D$^+$ + D fragmentation as a function of the pump-probe delay. A clear periodic oscillation is visible. A projection of this 2D spectrum to the delay axis is shown in Fig.~\ref{timeresolved}(d), indicating that the period of the NWP in the \Bstate state amounts on average to about 75~fs. The first maximum transition yield is reached at 45~fs. The corresponding spectrum for the photoelectron is shown in Fig.~\ref{timeresolved}(a).

In the 2D spectrum of Fig.~\ref{timeresolved}(e) we can also see that the KER depends on the pump-probe time delay. To quantify this observation we generated two projections of this 2D spectrum to show the ion yields as a function of time delay. This is presented in Fig.~\ref{timeresolved}(b) for a KER of 0.8~eV ($\pm$ 0.1~eV), and  in Fig.~\ref{timeresolved}(c), for a KER of 0.1~eV ($\pm$ 0.1~eV). The former channel dissociates sequentially: first, the NWP evolving on the \Bstate state is ionized to the bound \Xplusstate D$_2^+$ cation state via the absorption of 2 NIR photons. Around 10~fs later, at the outer wall of the \Xplusstate cation state, a transition to the repulsive 2p$\sigma_u$ D$_2^+$ cation state takes place, facilitated by the same NIR probe pulse, which is approximately 30~fs long. This leads to the dissociation with a rather high KER value of 0.8~eV. For smaller KER of around 0.1~eV, Fig.~\ref{timeresolved}(c) represents the evolution of the NWP on the \Bstate state and the higher electronic states, similar to the coincident pulse experiment, but with a lower XUV photon energy of 13.28~eV. We can see that on average the NWP needs about 55~fs to reach larger internuclear distances R enabling dissociative ionization. However, a significant portion of the wavepacket can reach such large R within even 20~fs. 

\begin{figure}
\includegraphics[trim={0cm 0cm 0cm 0cm},clip,width=1.0\columnwidth]{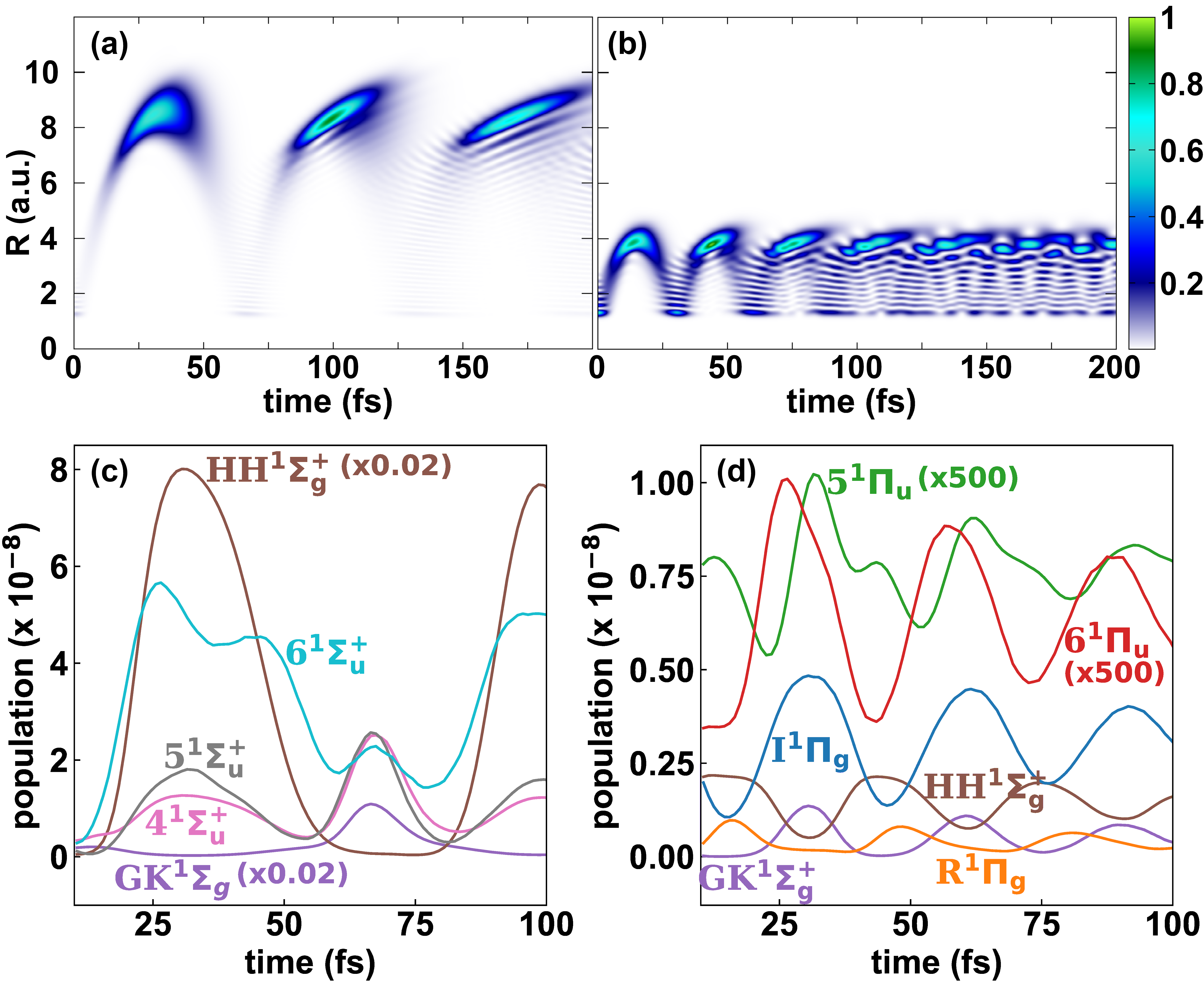}
\caption{\label{theoryfig}
Results from the present time-dependent Schr\"odinger equation calculations. The nuclear wavepacket amplitude (color scale) as a function of D--D internuclear distance R in atomic units (a.u.) and time delay relative to the instant the wavepacket is launched on the (a) \Bstate and (b) \Cstate electronic states. Selected electronic state populations relevant for the excitation pathways involving the (c) \Bstate and (d) \Cstate intermediate states (see text), as a function of time. The XUV and NIR photon energies are 13.95~eV and and 1.55~eV, respectively, and the pulse duration is 45~fs. In (c) and (d), the XUV + NIR excitation pulse is followed by a NIR probe pulse at time t, to simulate the time-dependence of electronic populations.}
\end{figure}

The results of TDSE calculations, which employ the higher XUV and NIR photon energies of 13.95~eV and 1.55~eV, respectively, the same as the coincident pulse experiment, are shown in Fig.~\ref{theoryfig}(a) and (c). We can see that the \HHstate, \FiveS, and \SixS states (brown, grey and cyan curves, respectively, in Fig.~\ref{theoryfig}(c)) are expected to be excited between 20 and 60~fs after the nuclear wavepacket is launched on the \Bstate state. This is the time for the XUV-excited NWP to reach the outer turning point of the \Bstate state (Fig.~\ref{theoryfig}(a)). We show this schematically in Fig.~\ref{scheme} (blue arrows), where the \SixS state is excited near the trailing edge of the pulse in the experiments. As the excited the neutral molecule continues to stretch, the D$^+$ + D KER spectrum is expected to reach a maximum value of KER = 240~meV extending all the way down to zero kinetic energy, which is consistent with the measured spectrum presented in Fig.~\ref{KERs}(a). The yield of the KER distribution plateaus at KER values of up to 120~meV. The total kinetic energy of the photoelectron and the dissociating molecule is conserved, so the expected electron energy E$_e$ has a plateau at E$_e$ = E$(h\nu) - E_{diss}(D_2^+)$ - KER = 18.6 - 18.158 - 0.12~eV = 0.322~eV. Given the NIR photon energy of 1.55~eV in our experiment and the energy difference between the intermediate neutral \SixS state and the final dissociative \SigmaUS cation state (or a degenerate continuum vibrational level of the \SigmaGS cation state), it is also conceivable for the neutral molecule to contract to just below R~$\le$~6~a.u. ($\le$~3.1~\AA) on the \SixS state where it still can be ionized with one NIR photon (as sketched in Fig.~\ref{scheme}, blue arrow). In this molecular contraction process the potential energy decreased from 17.05~eV to 16.85~eV and the electron is emitted with almost zero kinetic energy, while the KER has its maximum value of 240~meV.

It should be noted that the maximum sum photon energy of 18.6~eV, deposited into the molecular D$_2^+$ cation within the identified internuclear distance region of 6~a.u. to almost 9~a.u., enables the D$_2^+$ cation to dissociate directly on the \SigmaUS state or relax to the \SigmaGS state, before it finally dissociates. The necessary coupling for the latter process to occur is facilitated by the Coulomb potential of the emitted low energy photoelectron in the continuum via retroaction, as reported previously~\cite{serov_p_2014,waitz_electron_2016,heck_et_al_symmetry_nodate}. This retroaction effect results in an asymmetric photoelectron emission probability in the molecular frame, preferentially pointing along the charged fragment D$^+$ of the dissociating molecular cation D$_2^+$, which is on the order of $\le$ 15~\% for a mostly parallel transition with respect to the linear polarization direction of the ionizing photon pulse (see~\cite{heck_et_al_symmetry_nodate}). As this effect is rather small, and, as we will see, since parallel transitions are not favored in our sequential dissociation schemes discussed here, we will disregard its contribution in the analysis.

\begin{figure}
\begin{tabular}{cc}
\includegraphics[trim={6cm 2cm 4cm 1cm},clip,width=0.5\columnwidth]{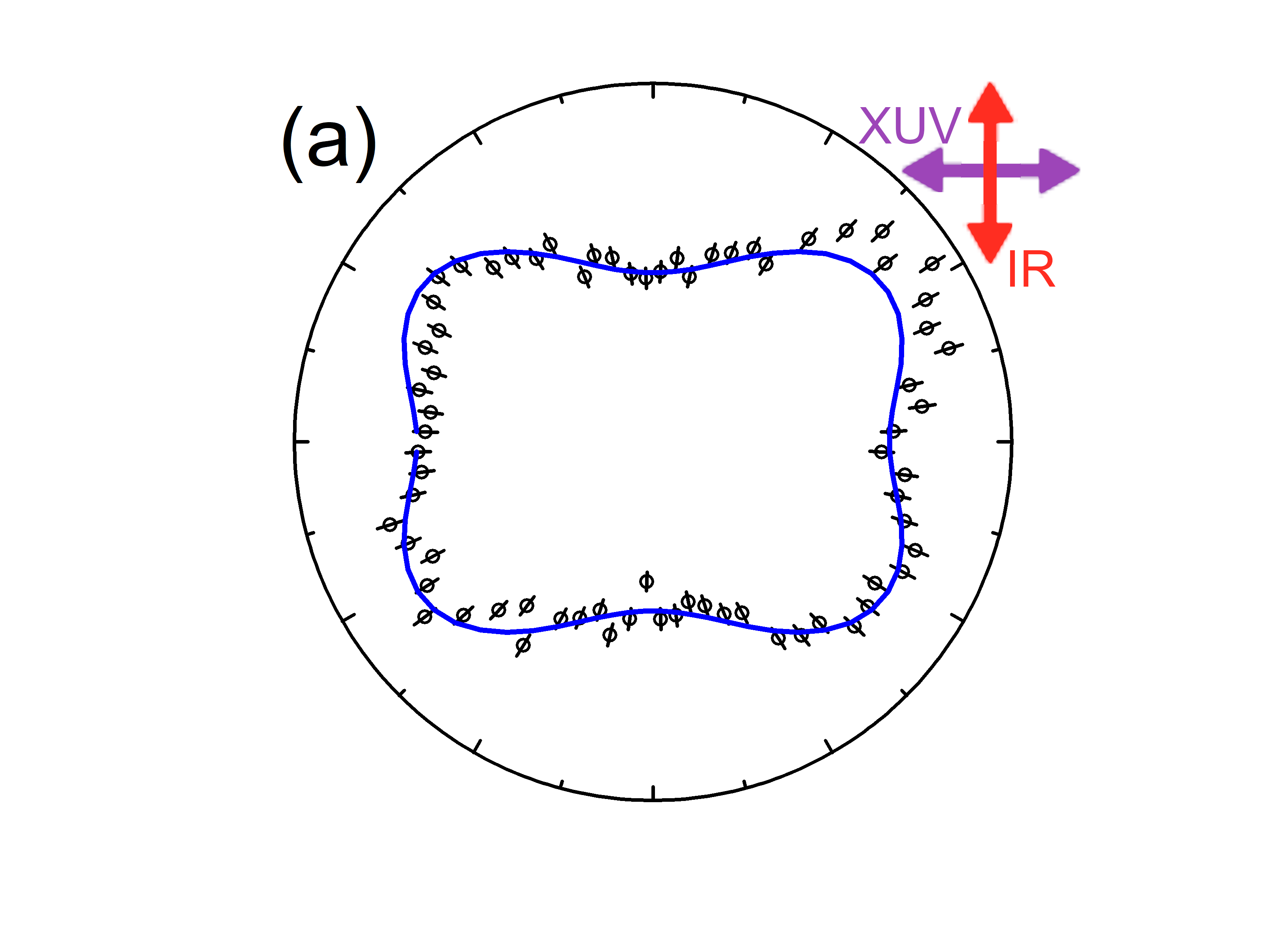} &
\includegraphics[trim={6cm 2cm 4cm 1cm},clip,width=0.5\columnwidth]{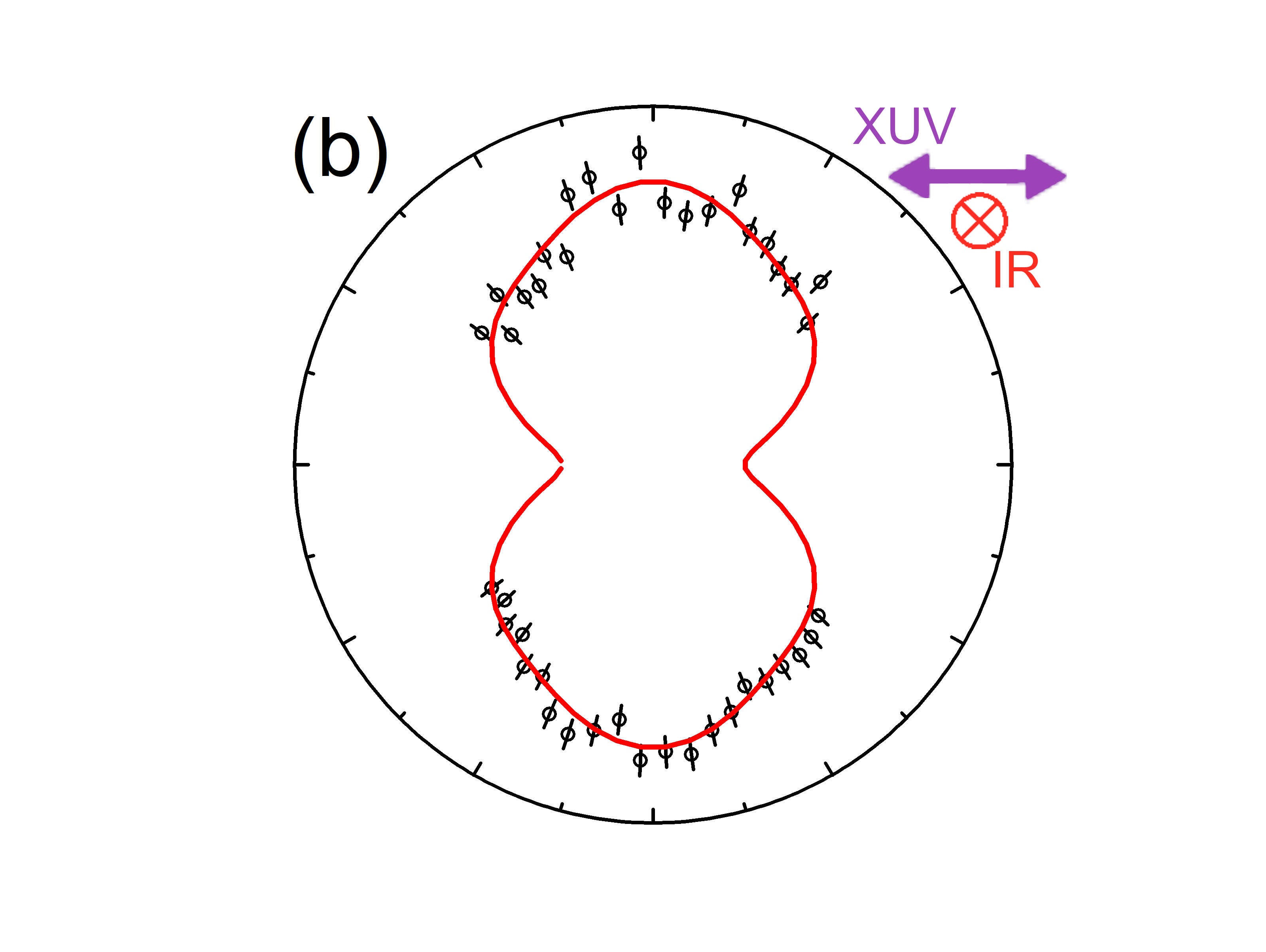}
\end{tabular}
\caption{\label{yz-angles}
(a) Measured (black open circles) D$^+$ azimuthal angular distribution for feature (III), and fit (blue line) using Equation~\ref{Ylmfit} for $L=0, 2, 4, 6$, $M=0$, $B_{20}=0.47 \pm 0.02$, $B_{40}=-0.19 \pm 0.01$ and $B_{60}=-0.07 \pm 0.02$. (b) Measured (black circles) D$^+$ angular distribution for feature (IV), and fit (red line) for $L=0, 2, 4, 6$, $M=0$, $B_{20}=-0.54 \pm 0.52$, $B_{40}=-0.06 \pm 0.37$ and $B_{60}=-0.10 \pm 0.15$. Only ions emitted in the range  The XUV polarization is in the horizontal direction (0$^\circ$ and 180$^\circ$) in each panel. The NIR polarization is vertical in panel (a) and orthogonal to the paper plane in panel (b). Only D$^+$ ions having momenta between 1~a.u. and 6~a.u. and emission angles within ($\theta_i \le \pm33^\circ$) of the relevant plane i are included in each plot.}
\end{figure}

In order to better understand the distinct preferred molecular orientations in the sequential photon absorption processes, i.e. the  cloverleaf-like ion momentum distribution (III) of Fig.~\ref{momenta}(a), we present the D$^+$ ion angular distribution in the plane parallel to both the NIR (y-axis) and the XUV polarization vector (z-axis) in Fig.~\ref{yz-angles}(a) for the same slicing condition ($\theta_i \le \pm33^\circ$). The plotted particle emission angle $\phi_\epsilon$ is an azimuthal angle ranging from -180$^\circ$ to 180$^\circ$ around the ordinate (photon propagation axis, x-axis) of this plane for D$^+$ ion momenta between 1~a.u. and 6~a.u., corresponding to ion energies between 0.01~eV and 0.12~eV. The $\phi_\epsilon$ angular distribution exhibits maxima at approximately $\pm$30$^\circ$ and $\pm$150$^\circ$ with respect to the XUV polarization direction and minima in each polarization direction, i.e. 0$^\circ$ and 180$^\circ$, the XUV polarization direction, as well as 90$^\circ$ and -90$^\circ$, the NIR polarization direction.

The photoion angular distributions resulting from a sequence of dipole transitions using linearly polarized fields are determined via selection rules, which place restrictions on the molecular orientations during photoabsorption. Such restrictions often cause nodal lines or planes in the particle emission patterns. Such nodal lines appear to be present in the momentum distribution of Fig.~\ref{momenta}(a). Like in Fig.~\ref{momenta}(a), Fig.~\ref{yz-angles}(a) shows minima in the D$^+$ angular distribution along the light polarization directions, where the finite acceptance angle of ($\theta_i \le \pm33^\circ$) determines the sharpness of these minima. This pattern is sensitive to the photoabsorption sequence involving parallel and perpendicular transitions leading to dissociative ionization. 

For multiphoton excitation of a diatomic molecule, the angular distribution $I(\theta_{\epsilon})$ of dissociating photo-fragments, defined by the polarization axis $\epsilon$ in the laboratory frame, can be expressed as an expansion in spherical harmonics of degree $L$ and order $M$:

\begin{equation}\label{Ylmfit}
\textrm I(\theta_{\epsilon},\phi_{\epsilon}) = \sum_{L,M} B_{LM} Y_{LM}(\theta_{\epsilon},\phi_{\epsilon}),
\end{equation}

where $\theta_{\epsilon}$ and $\phi_{\epsilon}$ are the polar and azimuthal angles, respectively, relative to $\epsilon$. 

In the present experiments, the XUV and NIR beams are cross-polarized, as indicated by the small pictograms in the upper right-hand corners in Fig.~\ref{momenta}. We first examine the XUV + 3 NIR sequence (III) in the y/z plane parallel to both the NIR and XUV polarization, and define the latter as the primary polarization direction $\epsilon$. For the crossed polarization and 4-photon photoionization, the D$^+$ fragment angular distribution can be most generally described using Equation \ref{Ylmfit}, allowing $L < 8$ and $-L \le M \le L$. However, we find that a satisfactory fit to the measured data can be achieved for $M=0$ and $L = 0, 2, 4, 6$. Such a weighted linear least squares fit to the measured angular distribution of feature (III) is displayed in Figure \ref{yz-angles}(a) as the blue line. Statistical experimental uncertainties are shown as error bars ($\pm$1 standard deviation), and the errors in the fit are determined using statistical bootstrapping~\cite{efronBetterBootstrapConfidence1987}. We find that the angular distribution is accurately parameterized by Equation \ref{Ylmfit} for $B_{20}=0.47 \pm 0.02$, $B_{40}=-0.19 \pm 0.01$ and $B_{60}=-0.07 \pm 0.02$.  

We now turn our attention to feature (IV) in the momentum distributions depicted in Fig.~\ref{momenta}(c) and (d), which both show D$^+$ ions being emitted along the light propagation direction (x-axis) with a discrete momentum centered around 3.5~a.u. This ion momentum corresponds to a kinetic energy peaking at around 40~meV, i.e. a D$^+$ + D KER of just $\thicksim$ 80~meV (see Fig.~\ref{KERs}(b)). The maximum value of the ion momentum distribution (p$_{x, D^+}$ $\le$ 4.0~a.u.) and kinetic energy release (KER $\le$ 220~meV) along the light direction in Fig.~\ref{momenta}(c) and (d) is perhaps slightly lower, but on the same order as the momenta composing feature (III) in the plane spanned by the NIR (y-axis) and XUV (z-axis) polarization vectors shown in Fig.~\ref{momenta}(a) and sketched in Fig.~\ref{momenta}(b). However, the momentum of ions in feature (IV) do not extend down to zero as in feature (III). Moreover, perpendicular transitions with respect to both polarization vectors are required for feature (IV) to produce D$^+$ emissions that are strongly aligned to the light propagation direction. Energy, angular momentum, and parity conservation suggests the following transition sequence (see Fig.~\ref{scheme}, green and red arrows):\\

(IV): 

\begin{align*}
\textrm{\GrndS}&\xrightarrow[res.]{XUV} \textrm{\Cstate}\\ &\xrightarrow[res.]{NIR} \textrm{\HHstate/\Istate/\GKstate}\\
&\xrightarrow[res.]{NIR} \textrm{\FiveP / \SixP}\\ &\xrightarrow[cont.]{NIR} \textrm{\SigmaUS} + \textrm{e}^-.
\end{align*}

From the sequence above, we understand feature (IV) is also generated via a 4-step dissociation process. In the first step, an XUV photon resonantly populates the neutral \Cstate state in a perpendicular transition at small internuclear distances within the Franck-Condon region (see Fig.~\ref{scheme}, green arrow), launching a vibrational nuclear wave packet on the excited electronic state. Calculations of the oscillator strengths (not shown here) suggest that an excitation of the first two vibrational levels of the 
\Dstate state is also possible, since the oscillator strengths are similar for each transition. Then, either a vibrational nuclear wave packet is launched and stretches up to 4.35~a.u. (2.3~\AA), before it returns to the Franck-Condon region, or two NIR quanta are immediately absorbed to resonantly populate the \HHstate/\Istate/\GKstate and the 
\FiveP and \SixP states with 17.05~eV in two resonant perpendicular transitions. According to our calculation, the contributions of the first three intermediate states oscillate in and out of phase with a period of around 30~fs. However, if the first three transitions all occur immediately in the Franck-Condon region, as sketched in Fig.~\ref{scheme} (red arrows), the \HHstate is the most likely intermediate state candidate to contribute to the pathway leading to the higher \FiveP and \SixP states. This pathway is supported by the TDSE calculations (Fig.~\ref{theoryfig}(d)), which also show that the \Istate and \GKstate intermediate states contribute significantly near the trailing edge of the NIR field, for t $>$ 20~fs.

On the 
\FiveP and \SixP states, a vibrational wave packet embarks and stretches towards larger internuclear distances. However, since the widths of the potential wells for the 
\FiveP and \SixP states are not as wide as that of the \Bstate state, the vibrational wave packet only extends to distances around R~=~5.6~a.u. (3.5~\AA), before it reaches its outer turning point at the potential barrier. At this turning point, by absorption of an additional NIR pulse (red arrow), a transition to the final \SigmaUS state of the D$_2^+$ cation in a perpendicular orientation of the molecule with respect to the NIR polarization is possible, producing the KER distribution to its prominent peak of 80~meV depicted in Fig.~\ref{KERs}(b). 

Two perpendicular transitions in the form of $\Sigma \rightarrow \Pi$ at the beginning and $\Pi \rightarrow \Sigma$ at the end of the dissociation sequence for both linearly cross-polarized polarization vectors of the XUV and NIR pulses can only be satisfied with the molecular axis being aligned along the photon propagation direction (x-axis), which is orthogonal to both XUV and NIR polarization directions of the two-color light pulses. This is with the measured D$^+$ ion momentum distributions in Fig.~\ref{momenta}(c) and (d) as well as the D$^+$ ion angular distribution presented in Fig.~\ref{yz-angles}(b). This ion angular distribution is plotted in the plane spanned by the photon propagation (x-axis) and the XUV polarization direction (z-axis) with an acceptance angle of $\theta_i \le \pm33^\circ$. Moreover, the D$^+$ ion momentum is restricted to be greater than 1~a.u. and smaller than 6~a.u.. The residual contribution from feature (III) is avoided by showing only ion emission angles from $\pm$45$^{\circ}$ to $\pm$135$^{\circ}$ in Fig.~\ref{yz-angles}(b). In isolating feature (IV) from feature (III), we limit the range for the angular distribution, leading to larger uncertainties for the fit using Equation~\ref{Ylmfit}. In this case we use $L=0, 2, 4, 6$, $M=0$, $B_{20}=-0.54 \pm 0.52$, $B_{40}=-0.06 \pm 0.37$ and $B_{60}=-0.10 \pm 0.15$ to achieve a plausible parameterization for this process, which is also guided by the respective momentum distributions shown in Fig.~\ref{momenta}(c)+(d) and sketched in Fig.~\ref{momenta}(b). 

We would like to stress that the D$^+$ ion emission pattern of feature (IV) along the light propagation direction distinguishes this fragmentation pathway from a dissociation sequence reported before (feature (V) in~\cite{sturm_mapping_2017}
), which resulted in a very similar KER but a breakup along the NIR polarization direction. This multi-step process involved an excitation to the \Dstatep state (and possibly to the \FiveP and \SixP states), either by the 9$^{th}$ harmonic in combination with two NIR photons or the 11$^{th}$ harmonic in those experiments. However, in that process, Rydberg states of the \state{\textrm{np}\pi}{1}{\Pi}{u} series could be populated that are known to predissociate through a non-adiabatic coupling into the H(1s) + H(3$p$) limit and which are ionized by one additional NIR photon according to the sequence:\\   

(V): 
\begin{align*}
\textrm{X}^{1}{\Sigma}_{g}&\xrightarrow[res.]{XUV + 2\cdot NIR} \textrm{D}'~^{1}\Pi_{u}(\textrm{4p}\pi) \xrightarrow[coupl.]{non-ad.}\textrm{H(1s)} + \textrm{H(3$p$)} \\
&\xrightarrow[cont.]{NIR} \textrm{H(1s)} + \textrm{H}^{+} + \textrm{e}^{-}.\\
\end{align*}

\section{\label{summary}Summary}

Neutral D$_2$ molecules were photoexcited and singly ionized close to threshold via two 4-photon absorption sequences. A linearly polarized 13.95~eV XUV pulse was combined with a 1.55~eV NIR pulse in a cross-polarized geometry, and the orientation of the molecular axis was detected with respect to the light polarization vectors with 3D fragment ion momentum imaging, revealing new structures in momentum space. This differential measurement enabled us to detect and analyze two sequential dissociation processes involving transitions occurring well-outside the ground state equilibrium geometries of H$_2$.

The dissociation process took place in a sequence involving excited electronic states of the neutral and ionized molecule. Dissociation fragments with very low KER ($\le$ 260~meV) were measured in two distinct processes distinguished by the fragment momentum distributions. The ion angular dependences in the laboratory frame correspond to D$_2$ molecules ionized with the molecular axis aligned either in the same plane including the XUV and NIR polarization directions, or orthogonal to that plane. These distinct breakup mechanisms were related to two possible fragmentation pathways, leading to the same final dissociating molecular D$_2^+$ cation state via different sequences of transitions between excited electronic states of the neutral D$_2$ molecule. Even in this fundamental and simple molecular system, the high abundance of electronic excited states creates some ambiguity in the correct assignment of the states involved in the dissociation sequences reported here. This concerns mostly the highly excited \SixS as well as 
the \FiveP and \SixP electronic states of the neutral D$_2$ molecule (see Fig.~\ref{scheme}). It is conceivable that there are states with the same symmetry within tenths of an eV of these states. However, the well-defined energy bandwidth and high momentum resolution of the experiment allowed two pathways to be identified as the most plausible reactions, with the aid of {\it ab initio} electronic structure and TDSE calculations to understand the propagation of the NWP for each pathway. 
In the first pathway [feature (III)] the XUV photon initiates a transition from the ground state of the neutral D$_2$ molecule to its electronically excited \Bstate state (see Fig.~\ref{scheme}, green arrow). A step-wise dissociation on the electronically excited neutral \Bstate state begins. A NWP is launched, where we measured the time it takes to reach large internuclear distances of up to 8.6~a.u. to be around 55~fs. This is followed by a dissociation on the \SixS state of D$_2$ after the resonant absorption of two NIR photons in sequence (see Fig.~\ref{scheme}, blue arrows). After absorbing an additional NIR photon, the dissociation sequence finishes on the \SigmaUS state of the molecular D$_2^+$ cation. This 4-step dissociation process is represented by a cloverleaf-like ion momentum distribution in the plane spanned by the XUV and NIR polarization vectors. The sharp nodal lines, corresponding to orientations of the molecular axis along either polarization vector, stem from the selection rules of the $\Sigma \rightarrow \Sigma \rightarrow \Sigma \rightarrow \Sigma \rightarrow \Sigma$ transition sequence, most visible in the D$^+$ ion momentum distribution [see Fig.~\ref{momenta}(a)]. 3D ion momentum imaging is shown here to be highly sensitive to the orientation of the dipole moments for specific sequential electronic transitions with respect to both XUV and NIR polarization vectors. This could be exploited as a detection scheme in future nonlinear spectroscopy experiments. 

The second pathway [feature (IV)] starts with a transition to the \Cstate state of the neutral D$_2$ molecule, initiated by the XUV pulse (see Fig.~\ref{scheme}, green arrow). Our measurement does not reveal if a vibrational wave-packet is launched and returns to the Franck-Condon region before the neutral molecule is further excited or if the neutral molecule immediately transitions to the 
\FiveP and \SixP states of D$_2$ by resonantly absorbing two NIR photons in combination with the XUV pulse in a vertical transition (see Fig.~\ref{scheme}, red arrows). 
The neutral molecule stretches on the high-lying intermediate 
\FiveP and \SixP states, however dissociation on these states is prevented by the low potential energy barrier. The dissociative \SigmaUS state of the D$_2^+$ cation is then populated in this stretched D$_2$ geometry by the third NIR photon (red arrow). The shape of the \FiveP and \SixP states, and the limited bandwidth of the XUV and NIR pulses, confine the internuclear distances over which dissociation on the \SigmaUS state is energetically accessible to approximately 4.5 to 5.5~a.u.  Accordingly, this multi-step dissociation process results in a KER distribution peak of $\sim 80$~meV. This sequence of $\Sigma \rightarrow \Pi \rightarrow \Pi/\Sigma \rightarrow \Pi \rightarrow \Sigma$ transitions in the cross-polarized XUV and NIR fields only fragments D$_2$ molecules that are orientated along the light propagation direction. This pathway could potentially be exploited for future time-resolved studies by photoelectron spectroscopy of fixed-in-space D$_2$ molecules. 
We hope this work inspires further highly detailed theoretical treatments, including predictions of ion angular distributions for few-photon dissociative photoionization. While this remains a challenging task, the H$_2$/D$_2$ molecule is an ideal test bed to extend existing theoretical methods in photoionization and dissociation.

\section*{Acknowledgments}

Work at Lawrence Berkeley National Laboratory was performed under the auspices of the U.S. Department of Energy under Contract No. DE-AC02-05CH11231, and was supported by the U.S. Department of Energy Office of Basic Energy Sciences, Division of Chemical Sciences, Biosciences and Geosciences. This research used resources of the National Energy Research Scientific Computing Center (NERSC), a U.S. Department of Energy Office of Science User Facilities under Contract No. DE-AC02-05CH11231. We are indebted to the RoentDek Company for long-term support with detector hardware and software.\\

\bibliography{D2DI-BBT}

\begin{thebibliography}{37}%
\makeatletter
\providecommand \@ifxundefined [1]{%
 \@ifx{#1\undefined}
}%
\providecommand \@ifnum [1]{%
 \ifnum #1\expandafter \@firstoftwo
 \else \expandafter \@secondoftwo
 \fi
}%
\providecommand \@ifx [1]{%
 \ifx #1\expandafter \@firstoftwo
 \else \expandafter \@secondoftwo
 \fi
}%
\providecommand \natexlab [1]{#1}%
\providecommand \enquote  [1]{``#1''}%
\providecommand \bibnamefont  [1]{#1}%
\providecommand \bibfnamefont [1]{#1}%
\providecommand \citenamefont [1]{#1}%
\providecommand \href@noop [0]{\@secondoftwo}%
\providecommand \href [0]{\begingroup \@sanitize@url \@href}%
\providecommand \@href[1]{\@@startlink{#1}\@@href}%
\providecommand \@@href[1]{\endgroup#1\@@endlink}%
\providecommand \@sanitize@url [0]{\catcode `\\12\catcode `\$12\catcode
  `\&12\catcode `\#12\catcode `\^12\catcode `\_12\catcode `\%12\relax}%
\providecommand \@@startlink[1]{}%
\providecommand \@@endlink[0]{}%
\providecommand \url  [0]{\begingroup\@sanitize@url \@url }%
\providecommand \@url [1]{\endgroup\@href {#1}{\urlprefix }}%
\providecommand \urlprefix  [0]{URL }%
\providecommand \Eprint [0]{\href }%
\providecommand \doibase [0]{http://dx.doi.org/}%
\providecommand \selectlanguage [0]{\@gobble}%
\providecommand \bibinfo  [0]{\@secondoftwo}%
\providecommand \bibfield  [0]{\@secondoftwo}%
\providecommand \translation [1]{[#1]}%
\providecommand \BibitemOpen [0]{}%
\providecommand \bibitemStop [0]{}%
\providecommand \bibitemNoStop [0]{.\EOS\space}%
\providecommand \EOS [0]{\spacefactor3000\relax}%
\providecommand \BibitemShut  [1]{\csname bibitem#1\endcsname}%
\let\auto@bib@innerbib\@empty
\bibitem [{\citenamefont {Hemminger}\ \emph {et~al.}(2015)\citenamefont
  {Hemminger}, \citenamefont {Sarrao}, \citenamefont {Crabtree}, \citenamefont
  {Flemming},\ and\ \citenamefont {Ratner}}]{hemminger_challenges_2015}%
  \BibitemOpen
  \bibfield  {author} {\bibinfo {author} {\bibfnamefont {J.~C.}\ \bibnamefont
  {Hemminger}}, \bibinfo {author} {\bibfnamefont {J.}~\bibnamefont {Sarrao}},
  \bibinfo {author} {\bibfnamefont {G.}~\bibnamefont {Crabtree}}, \bibinfo
  {author} {\bibfnamefont {G.}~\bibnamefont {Flemming}}, \ and\ \bibinfo
  {author} {\bibfnamefont {M.}~\bibnamefont {Ratner}},\ }\href@noop {} {\emph
  {\bibinfo {title} {Challenges at the {{Frontiers}} of {{Matter}} and
  {{Energy}}: {{Transformative Opportunities}} for {{Discovery Science}}}}},\
  \bibinfo {type} {Tech. Rep.}\ \bibinfo {number} {1283188}\ (\bibinfo {year}
  {2015})\BibitemShut {NoStop}%
\bibitem [{\citenamefont {Sandhu}\ \emph {et~al.}(2008)\citenamefont {Sandhu},
  \citenamefont {Gagnon}, \citenamefont {Santra}, \citenamefont {Sharma},
  \citenamefont {Li}, \citenamefont {Ho}, \citenamefont {Ranitovic},
  \citenamefont {Cocke}, \citenamefont {Murnane},\ and\ \citenamefont
  {Kapteyn}}]{sandhu_observing_2008}%
  \BibitemOpen
  \bibfield  {author} {\bibinfo {author} {\bibfnamefont {A.~S.}\ \bibnamefont
  {Sandhu}}, \bibinfo {author} {\bibfnamefont {E.}~\bibnamefont {Gagnon}},
  \bibinfo {author} {\bibfnamefont {R.}~\bibnamefont {Santra}}, \bibinfo
  {author} {\bibfnamefont {V.}~\bibnamefont {Sharma}}, \bibinfo {author}
  {\bibfnamefont {W.}~\bibnamefont {Li}}, \bibinfo {author} {\bibfnamefont
  {P.}~\bibnamefont {Ho}}, \bibinfo {author} {\bibfnamefont {P.}~\bibnamefont
  {Ranitovic}}, \bibinfo {author} {\bibfnamefont {C.~L.}\ \bibnamefont
  {Cocke}}, \bibinfo {author} {\bibfnamefont {M.~M.}\ \bibnamefont {Murnane}},
  \ and\ \bibinfo {author} {\bibfnamefont {H.~C.}\ \bibnamefont {Kapteyn}},\
  }\href@noop {} {\bibfield  {journal} {\bibinfo  {journal} {Science}\ }\textbf
  {\bibinfo {volume} {322}},\ \bibinfo {pages} {1081} (\bibinfo {year}
  {2008})}\BibitemShut {NoStop}%
\bibitem [{\citenamefont {Zhou}\ \emph {et~al.}(2012)\citenamefont {Zhou},
  \citenamefont {Ranitovic}, \citenamefont {Hogle}, \citenamefont {Eland},
  \citenamefont {Kapteyn},\ and\ \citenamefont {Murnane}}]{zhou_probing_2012}%
  \BibitemOpen
  \bibfield  {author} {\bibinfo {author} {\bibfnamefont {X.}~\bibnamefont
  {Zhou}}, \bibinfo {author} {\bibfnamefont {P.}~\bibnamefont {Ranitovic}},
  \bibinfo {author} {\bibfnamefont {C.~W.}\ \bibnamefont {Hogle}}, \bibinfo
  {author} {\bibfnamefont {J.~H.~D.}\ \bibnamefont {Eland}}, \bibinfo {author}
  {\bibfnamefont {H.~C.}\ \bibnamefont {Kapteyn}}, \ and\ \bibinfo {author}
  {\bibfnamefont {M.~M.}\ \bibnamefont {Murnane}},\ }\href@noop {} {\bibfield
  {journal} {\bibinfo  {journal} {Nat. Phys.}\ }\textbf {\bibinfo {volume}
  {8}},\ \bibinfo {pages} {232} (\bibinfo {year} {2012})}\BibitemShut {NoStop}%
\bibitem [{\citenamefont {Champenois}\ \emph {et~al.}(2016)\citenamefont
  {Champenois}, \citenamefont {Shivaram}, \citenamefont {Wright}, \citenamefont
  {Yang}, \citenamefont {Belkacem},\ and\ \citenamefont
  {Cryan}}]{champenois_involvement_2016}%
  \BibitemOpen
  \bibfield  {author} {\bibinfo {author} {\bibfnamefont {E.~G.}\ \bibnamefont
  {Champenois}}, \bibinfo {author} {\bibfnamefont {N.~H.}\ \bibnamefont
  {Shivaram}}, \bibinfo {author} {\bibfnamefont {T.~W.}\ \bibnamefont
  {Wright}}, \bibinfo {author} {\bibfnamefont {C.-S.}\ \bibnamefont {Yang}},
  \bibinfo {author} {\bibfnamefont {A.}~\bibnamefont {Belkacem}}, \ and\
  \bibinfo {author} {\bibfnamefont {J.~P.}\ \bibnamefont {Cryan}},\ }\href@noop
  {} {\bibfield  {journal} {\bibinfo  {journal} {The Journal of Chemical
  Physics}\ }\textbf {\bibinfo {volume} {144}},\ \bibinfo {pages} {014303}
  (\bibinfo {year} {2016})}\BibitemShut {NoStop}%
\bibitem [{\citenamefont {L{\'e}pine}\ \emph {et~al.}(2013)\citenamefont
  {L{\'e}pine}, \citenamefont {Sansone},\ and\ \citenamefont
  {Vrakking}}]{lepine_molecular_2013}%
  \BibitemOpen
  \bibfield  {author} {\bibinfo {author} {\bibfnamefont {F.}~\bibnamefont
  {L{\'e}pine}}, \bibinfo {author} {\bibfnamefont {G.}~\bibnamefont {Sansone}},
  \ and\ \bibinfo {author} {\bibfnamefont {M.~J.}\ \bibnamefont {Vrakking}},\
  }\href@noop {} {\bibfield  {journal} {\bibinfo  {journal} {Chem. Phys.
  Lett.}\ }\textbf {\bibinfo {volume} {578}},\ \bibinfo {pages} {1} (\bibinfo
  {year} {2013})}\BibitemShut {NoStop}%
\bibitem [{\citenamefont {Cao}\ \emph {et~al.}(2015)\citenamefont {Cao},
  \citenamefont {Laurent}, \citenamefont {{Ben-Itzhak}},\ and\ \citenamefont
  {Cocke}}]{cao_identification_2015}%
  \BibitemOpen
  \bibfield  {author} {\bibinfo {author} {\bibfnamefont {W.}~\bibnamefont
  {Cao}}, \bibinfo {author} {\bibfnamefont {G.}~\bibnamefont {Laurent}},
  \bibinfo {author} {\bibfnamefont {I.}~\bibnamefont {{Ben-Itzhak}}}, \ and\
  \bibinfo {author} {\bibfnamefont {C.~L.}\ \bibnamefont {Cocke}},\ }\href@noop
  {} {\bibfield  {journal} {\bibinfo  {journal} {Phys. Rev. Lett.}\ }\textbf
  {\bibinfo {volume} {114}} (\bibinfo {year} {2015})}\BibitemShut {NoStop}%
\bibitem [{\citenamefont {Sturm}\ \emph {et~al.}(2017)\citenamefont {Sturm},
  \citenamefont {Tong}, \citenamefont {Palacios}, \citenamefont {Wright},
  \citenamefont {Zalyubovskaya}, \citenamefont {Ray}, \citenamefont {Shivaram},
  \citenamefont {Mart{\'i}n}, \citenamefont {Belkacem}, \citenamefont
  {Ranitovic},\ and\ \citenamefont {Weber}}]{sturm_mapping_2017}%
  \BibitemOpen
  \bibfield  {author} {\bibinfo {author} {\bibfnamefont {F.~P.}\ \bibnamefont
  {Sturm}}, \bibinfo {author} {\bibfnamefont {X.~M.}\ \bibnamefont {Tong}},
  \bibinfo {author} {\bibfnamefont {A.}~\bibnamefont {Palacios}}, \bibinfo
  {author} {\bibfnamefont {T.~W.}\ \bibnamefont {Wright}}, \bibinfo {author}
  {\bibfnamefont {I.}~\bibnamefont {Zalyubovskaya}}, \bibinfo {author}
  {\bibfnamefont {D.}~\bibnamefont {Ray}}, \bibinfo {author} {\bibfnamefont
  {N.}~\bibnamefont {Shivaram}}, \bibinfo {author} {\bibfnamefont
  {F.}~\bibnamefont {Mart{\'i}n}}, \bibinfo {author} {\bibfnamefont
  {A.}~\bibnamefont {Belkacem}}, \bibinfo {author} {\bibfnamefont
  {P.}~\bibnamefont {Ranitovic}}, \ and\ \bibinfo {author} {\bibfnamefont
  {T.}~\bibnamefont {Weber}},\ }\href@noop {} {\bibfield  {journal} {\bibinfo
  {journal} {Phys. Rev. A}\ }\textbf {\bibinfo {volume} {95}} (\bibinfo {year}
  {2017})}\BibitemShut {NoStop}%
\bibitem [{\citenamefont {Zewail}(2000)}]{zewail_femtochemistry_2000}%
  \BibitemOpen
  \bibfield  {author} {\bibinfo {author} {\bibfnamefont {A.~H.}\ \bibnamefont
  {Zewail}},\ }\href@noop {} {\bibfield  {journal} {\bibinfo  {journal} {J.
  Phys. Chem. A}\ }\textbf {\bibinfo {volume} {104}},\ \bibinfo {pages} {5660}
  (\bibinfo {year} {2000})}\BibitemShut {NoStop}%
\bibitem [{\citenamefont {Sarkisov}\ and\ \citenamefont
  {Umanskii}(2001)}]{sarkisov_femtochemistry_2001}%
  \BibitemOpen
  \bibfield  {author} {\bibinfo {author} {\bibfnamefont {O.~M.}\ \bibnamefont
  {Sarkisov}}\ and\ \bibinfo {author} {\bibfnamefont {S.~Y.}\ \bibnamefont
  {Umanskii}},\ }\href@noop {} {\bibfield  {journal} {\bibinfo  {journal}
  {Russ. Chem. Rev.}\ }\textbf {\bibinfo {volume} {70}},\ \bibinfo {pages}
  {449} (\bibinfo {year} {2001})}\BibitemShut {NoStop}%
\bibitem [{\citenamefont {Calegari}\ \emph {et~al.}(2016)\citenamefont
  {Calegari}, \citenamefont {Sansone}, \citenamefont {Stagira}, \citenamefont
  {Vozzi},\ and\ \citenamefont {Nisoli}}]{calegari_advances_2016}%
  \BibitemOpen
  \bibfield  {author} {\bibinfo {author} {\bibfnamefont {F.}~\bibnamefont
  {Calegari}}, \bibinfo {author} {\bibfnamefont {G.}~\bibnamefont {Sansone}},
  \bibinfo {author} {\bibfnamefont {S.}~\bibnamefont {Stagira}}, \bibinfo
  {author} {\bibfnamefont {C.}~\bibnamefont {Vozzi}}, \ and\ \bibinfo {author}
  {\bibfnamefont {M.}~\bibnamefont {Nisoli}},\ }\href@noop {} {\bibfield
  {journal} {\bibinfo  {journal} {J. Phys. B At. Mol. Opt. Phys.}\ }\textbf
  {\bibinfo {volume} {49}},\ \bibinfo {pages} {062001} (\bibinfo {year}
  {2016})}\BibitemShut {NoStop}%
\bibitem [{\citenamefont {Drescher}\ \emph {et~al.}(2016)\citenamefont
  {Drescher}, \citenamefont {Galbraith}, \citenamefont {Reitsma}, \citenamefont
  {Dura}, \citenamefont {Zhavoronkov}, \citenamefont {Patchkovskii},
  \citenamefont {Vrakking},\ and\ \citenamefont
  {Mikosch}}]{drescher_communication_2016}%
  \BibitemOpen
  \bibfield  {author} {\bibinfo {author} {\bibfnamefont {L.}~\bibnamefont
  {Drescher}}, \bibinfo {author} {\bibfnamefont {M.~C.~E.}\ \bibnamefont
  {Galbraith}}, \bibinfo {author} {\bibfnamefont {G.}~\bibnamefont {Reitsma}},
  \bibinfo {author} {\bibfnamefont {J.}~\bibnamefont {Dura}}, \bibinfo {author}
  {\bibfnamefont {N.}~\bibnamefont {Zhavoronkov}}, \bibinfo {author}
  {\bibfnamefont {S.}~\bibnamefont {Patchkovskii}}, \bibinfo {author}
  {\bibfnamefont {M.~J.~J.}\ \bibnamefont {Vrakking}}, \ and\ \bibinfo {author}
  {\bibfnamefont {J.}~\bibnamefont {Mikosch}},\ }\href@noop {} {\bibfield
  {journal} {\bibinfo  {journal} {J. Chem. Phys.}\ }\textbf {\bibinfo {volume}
  {145}},\ \bibinfo {pages} {011101} (\bibinfo {year} {2016})}\BibitemShut
  {NoStop}%
\bibitem [{\citenamefont {Johnsson}\ \emph {et~al.}(2007)\citenamefont
  {Johnsson}, \citenamefont {Mauritsson}, \citenamefont {Remetter},
  \citenamefont {L'Huillier},\ and\ \citenamefont
  {Schafer}}]{johnsson_attosecond_2007}%
  \BibitemOpen
  \bibfield  {author} {\bibinfo {author} {\bibfnamefont {P.}~\bibnamefont
  {Johnsson}}, \bibinfo {author} {\bibfnamefont {J.}~\bibnamefont
  {Mauritsson}}, \bibinfo {author} {\bibfnamefont {T.}~\bibnamefont
  {Remetter}}, \bibinfo {author} {\bibfnamefont {A.}~\bibnamefont
  {L'Huillier}}, \ and\ \bibinfo {author} {\bibfnamefont {K.~J.}\ \bibnamefont
  {Schafer}},\ }\href@noop {} {\bibfield  {journal} {\bibinfo  {journal} {Phys.
  Rev. Lett.}\ }\textbf {\bibinfo {volume} {99}} (\bibinfo {year}
  {2007})}\BibitemShut {NoStop}%
\bibitem [{\citenamefont {Ranitovic}\ \emph {et~al.}(2014)\citenamefont
  {Ranitovic}, \citenamefont {Hogle}, \citenamefont {Riviere}, \citenamefont
  {Palacios}, \citenamefont {Tong}, \citenamefont {Toshima}, \citenamefont
  {{Gonzalez-Castrillo}}, \citenamefont {Martin}, \citenamefont {Martin},
  \citenamefont {Murnane},\ and\ \citenamefont
  {Kapteyn}}]{ranitovic_attosecond_2014}%
  \BibitemOpen
  \bibfield  {author} {\bibinfo {author} {\bibfnamefont {P.}~\bibnamefont
  {Ranitovic}}, \bibinfo {author} {\bibfnamefont {C.~W.}\ \bibnamefont
  {Hogle}}, \bibinfo {author} {\bibfnamefont {P.}~\bibnamefont {Riviere}},
  \bibinfo {author} {\bibfnamefont {A.}~\bibnamefont {Palacios}}, \bibinfo
  {author} {\bibfnamefont {X.-M.}\ \bibnamefont {Tong}}, \bibinfo {author}
  {\bibfnamefont {N.}~\bibnamefont {Toshima}}, \bibinfo {author} {\bibfnamefont
  {A.}~\bibnamefont {{Gonzalez-Castrillo}}}, \bibinfo {author} {\bibfnamefont
  {L.}~\bibnamefont {Martin}}, \bibinfo {author} {\bibfnamefont
  {F.}~\bibnamefont {Martin}}, \bibinfo {author} {\bibfnamefont {M.~M.}\
  \bibnamefont {Murnane}}, \ and\ \bibinfo {author} {\bibfnamefont
  {H.}~\bibnamefont {Kapteyn}},\ }\href@noop {} {\bibfield  {journal} {\bibinfo
   {journal} {Proc. Natl. Acad. Sci.}\ }\textbf {\bibinfo {volume} {111}},\
  \bibinfo {pages} {912} (\bibinfo {year} {2014})}\BibitemShut {NoStop}%
\bibitem [{\citenamefont {Larsen}\ \emph {et~al.}(2020)\citenamefont {Larsen},
  \citenamefont {Lucchese}, \citenamefont {Slaughter},\ and\ \citenamefont
  {Weber}}]{Larsen}%
  \BibitemOpen
  \bibfield  {author} {\bibinfo {author} {\bibfnamefont {K.~A.}\ \bibnamefont
  {Larsen}}, \bibinfo {author} {\bibfnamefont {R.~R.}\ \bibnamefont
  {Lucchese}}, \bibinfo {author} {\bibfnamefont {D.~S.}\ \bibnamefont
  {Slaughter}}, \ and\ \bibinfo {author} {\bibfnamefont {T.}~\bibnamefont
  {Weber}},\ }\href@noop {} {\bibfield  {journal} {\bibinfo  {journal} {The
  Journal of Chemical Physics}\ }\textbf {\bibinfo {volume} {153}},\ \bibinfo
  {pages} {021103} (\bibinfo {year} {2020})}\BibitemShut {NoStop}%
\bibitem [{\citenamefont {Fantz}\ and\ \citenamefont
  {W{\"u}nderlich}(2006)}]{fantz_franckcondon_2006}%
  \BibitemOpen
  \bibfield  {author} {\bibinfo {author} {\bibfnamefont {U.}~\bibnamefont
  {Fantz}}\ and\ \bibinfo {author} {\bibfnamefont {D.}~\bibnamefont
  {W{\"u}nderlich}},\ }\href@noop {} {\bibfield  {journal} {\bibinfo  {journal}
  {At. Data Nucl. Data Tables}\ }\textbf {\bibinfo {volume} {92}},\ \bibinfo
  {pages} {853} (\bibinfo {year} {2006})}\BibitemShut {NoStop}%
\bibitem [{\citenamefont {Sturm}\ \emph {et~al.}(2016)\citenamefont {Sturm},
  \citenamefont {Wright}, \citenamefont {Ray}, \citenamefont {Zalyubovskaya},
  \citenamefont {Shivaram}, \citenamefont {Slaughter}, \citenamefont
  {Ranitovic}, \citenamefont {Belkacem},\ and\ \citenamefont
  {Weber}}]{sturm_time_2016}%
  \BibitemOpen
  \bibfield  {author} {\bibinfo {author} {\bibfnamefont {F.~P.}\ \bibnamefont
  {Sturm}}, \bibinfo {author} {\bibfnamefont {T.~W.}\ \bibnamefont {Wright}},
  \bibinfo {author} {\bibfnamefont {D.}~\bibnamefont {Ray}}, \bibinfo {author}
  {\bibfnamefont {I.}~\bibnamefont {Zalyubovskaya}}, \bibinfo {author}
  {\bibfnamefont {N.}~\bibnamefont {Shivaram}}, \bibinfo {author}
  {\bibfnamefont {D.~S.}\ \bibnamefont {Slaughter}}, \bibinfo {author}
  {\bibfnamefont {P.}~\bibnamefont {Ranitovic}}, \bibinfo {author}
  {\bibfnamefont {A.}~\bibnamefont {Belkacem}}, \ and\ \bibinfo {author}
  {\bibfnamefont {T.}~\bibnamefont {Weber}},\ }\href@noop {} {\bibfield
  {journal} {\bibinfo  {journal} {Rev. Sci. Instrum.}\ }\textbf {\bibinfo
  {volume} {87}},\ \bibinfo {pages} {063110} (\bibinfo {year}
  {2016})}\BibitemShut {NoStop}%
\bibitem [{\citenamefont {D{\"o}rner}\ \emph {et~al.}(2000)\citenamefont
  {D{\"o}rner}, \citenamefont {Mergel}, \citenamefont {Jagutzki}, \citenamefont
  {Spielberger}, \citenamefont {Ullrich}, \citenamefont {Moshammer},\ and\
  \citenamefont {{Schmidt-B{\"o}cking}}}]{dorner_cold_2000}%
  \BibitemOpen
  \bibfield  {author} {\bibinfo {author} {\bibfnamefont {R.}~\bibnamefont
  {D{\"o}rner}}, \bibinfo {author} {\bibfnamefont {V.}~\bibnamefont {Mergel}},
  \bibinfo {author} {\bibfnamefont {O.}~\bibnamefont {Jagutzki}}, \bibinfo
  {author} {\bibfnamefont {L.}~\bibnamefont {Spielberger}}, \bibinfo {author}
  {\bibfnamefont {J.}~\bibnamefont {Ullrich}}, \bibinfo {author} {\bibfnamefont
  {R.}~\bibnamefont {Moshammer}}, \ and\ \bibinfo {author} {\bibfnamefont
  {H.}~\bibnamefont {{Schmidt-B{\"o}cking}}},\ }\href@noop {} {\bibfield
  {journal} {\bibinfo  {journal} {Phys. Rep.}\ }\textbf {\bibinfo {volume}
  {330}},\ \bibinfo {pages} {95} (\bibinfo {year} {2000})}\BibitemShut
  {NoStop}%
\bibitem [{\citenamefont {Ullrich}\ \emph {et~al.}(2003)\citenamefont
  {Ullrich}, \citenamefont {Moshammer}, \citenamefont {Dorn}, \citenamefont
  {D{\"o}rner}, \citenamefont {Schmidt},\ and\ \citenamefont
  {{Schmidt-B{\"o}cking}}}]{ullrich_recoil-ion_2003}%
  \BibitemOpen
  \bibfield  {author} {\bibinfo {author} {\bibfnamefont {J.}~\bibnamefont
  {Ullrich}}, \bibinfo {author} {\bibfnamefont {R.}~\bibnamefont {Moshammer}},
  \bibinfo {author} {\bibfnamefont {A.}~\bibnamefont {Dorn}}, \bibinfo {author}
  {\bibfnamefont {R.}~\bibnamefont {D{\"o}rner}}, \bibinfo {author}
  {\bibfnamefont {L.~P.~H.}\ \bibnamefont {Schmidt}}, \ and\ \bibinfo {author}
  {\bibfnamefont {H.}~\bibnamefont {{Schmidt-B{\"o}cking}}},\ }\href@noop {}
  {\bibfield  {journal} {\bibinfo  {journal} {Rep. Prog. Phys.}\ }\textbf
  {\bibinfo {volume} {66}},\ \bibinfo {pages} {1463} (\bibinfo {year}
  {2003})}\BibitemShut {NoStop}%
\bibitem [{\citenamefont {{Roentdek}}()}]{roentdek_mcp_nodate}%
  \BibitemOpen
  \bibfield  {author} {\bibinfo {author} {\bibnamefont {{Roentdek}}},\
  }\href@noop {} {\enquote {\bibinfo {title} {{{MCP Delay Line Detector
  Manual}}},}\ }\BibitemShut {NoStop}%
\bibitem [{\citenamefont {Ranitovic}\ \emph {et~al.}(2018)\citenamefont
  {Ranitovic}, \citenamefont {Sturm}, \citenamefont {Tong}, \citenamefont
  {Wright}, \citenamefont {Ray}, \citenamefont {Zalyubovskya}, \citenamefont
  {Shivaram}, \citenamefont {Belkacem}, \citenamefont {Slaughter},\ and\
  \citenamefont {Weber}}]{ranitovic_attosecond_2018}%
  \BibitemOpen
  \bibfield  {author} {\bibinfo {author} {\bibfnamefont {P.}~\bibnamefont
  {Ranitovic}}, \bibinfo {author} {\bibfnamefont {F.~P.}\ \bibnamefont
  {Sturm}}, \bibinfo {author} {\bibfnamefont {X.~M.}\ \bibnamefont {Tong}},
  \bibinfo {author} {\bibfnamefont {T.~W.}\ \bibnamefont {Wright}}, \bibinfo
  {author} {\bibfnamefont {D.}~\bibnamefont {Ray}}, \bibinfo {author}
  {\bibfnamefont {I.}~\bibnamefont {Zalyubovskya}}, \bibinfo {author}
  {\bibfnamefont {N.}~\bibnamefont {Shivaram}}, \bibinfo {author}
  {\bibfnamefont {A.}~\bibnamefont {Belkacem}}, \bibinfo {author}
  {\bibfnamefont {D.~S.}\ \bibnamefont {Slaughter}}, \ and\ \bibinfo {author}
  {\bibfnamefont {T.}~\bibnamefont {Weber}},\ }\href@noop {} {\bibfield
  {journal} {\bibinfo  {journal} {Phys. Rev. A}\ }\textbf {\bibinfo {volume}
  {98}},\ \bibinfo {pages} {013410} (\bibinfo {year} {2018})}\BibitemShut
  {NoStop}%
\bibitem [{\citenamefont {Werner}\ \emph {et~al.}(2012)\citenamefont {Werner},
  \citenamefont {Knowles}, \citenamefont {Knizia}, \citenamefont {Manby},\ and\
  \citenamefont {Sch\"utz}}]{werner_molpro_2012}%
  \BibitemOpen
  \bibfield  {author} {\bibinfo {author} {\bibfnamefont {H.-J.}\ \bibnamefont
  {Werner}}, \bibinfo {author} {\bibfnamefont {P.~J.}\ \bibnamefont {Knowles}},
  \bibinfo {author} {\bibfnamefont {G.}~\bibnamefont {Knizia}}, \bibinfo
  {author} {\bibfnamefont {F.~R.}\ \bibnamefont {Manby}}, \ and\ \bibinfo
  {author} {\bibfnamefont {M.}~\bibnamefont {Sch\"utz}},\ }\href {\doibase
  10.1002/wcms.82} {\bibfield  {journal} {\bibinfo  {journal} {WIREs
  Computational Molecular Science}\ }\textbf {\bibinfo {volume} {2}},\ \bibinfo
  {pages} {242} (\bibinfo {year} {2012})}\BibitemShut {NoStop}%
\bibitem [{\citenamefont {Werner}\ \emph {et~al.}(2015)\citenamefont {Werner},
  \citenamefont {Knowles}, \citenamefont {Knizia}, \citenamefont {Manby},
  \citenamefont {Sch\"utz}, \citenamefont {Celani}, \citenamefont {Korona},
  \citenamefont {Lindh}, \citenamefont {Mitrushenkov}, \citenamefont {Rauhut},
  \citenamefont {Shamasundar}, \citenamefont {Adler}, \citenamefont {Amos},
  \citenamefont {Bernhardsson}, \citenamefont {Berning}, \citenamefont
  {Cooper}, \citenamefont {Deegan}, \citenamefont {Dobbyn}, \citenamefont
  {Eckert}, \citenamefont {Goll}, \citenamefont {Hampel}, \citenamefont
  {Hesselmann}, \citenamefont {Hetzer}, \citenamefont {Hrenar}, \citenamefont
  {Jansen}, \citenamefont {K\"oppl}, \citenamefont {Liu}, \citenamefont
  {Lloyd}, \citenamefont {Mata}, \citenamefont {May}, \citenamefont
  {McNicholas}, \citenamefont {Meyer}, \citenamefont {Mura}, \citenamefont
  {Nicklass}, \citenamefont {O'Neill}, \citenamefont {Palmieri}, \citenamefont
  {Peng}, \citenamefont {Pfl\"uger}, \citenamefont {Pitzer}, \citenamefont
  {Reiher}, \citenamefont {Shiozaki}, \citenamefont {Stoll}, \citenamefont
  {Stone}, \citenamefont {Tarroni}, \citenamefont {Thorsteinsson},\ and\
  \citenamefont {Wang}}]{werner_molpro_2015}%
  \BibitemOpen
  \bibfield  {author} {\bibinfo {author} {\bibfnamefont {H.-J.}\ \bibnamefont
  {Werner}}, \bibinfo {author} {\bibfnamefont {P.~J.}\ \bibnamefont {Knowles}},
  \bibinfo {author} {\bibfnamefont {G.}~\bibnamefont {Knizia}}, \bibinfo
  {author} {\bibfnamefont {F.~R.}\ \bibnamefont {Manby}}, \bibinfo {author}
  {\bibfnamefont {M.}~\bibnamefont {Sch\"utz}}, \bibinfo {author}
  {\bibfnamefont {P.}~\bibnamefont {Celani}}, \bibinfo {author} {\bibfnamefont
  {T.}~\bibnamefont {Korona}}, \bibinfo {author} {\bibfnamefont
  {R.}~\bibnamefont {Lindh}}, \bibinfo {author} {\bibfnamefont
  {A.}~\bibnamefont {Mitrushenkov}}, \bibinfo {author} {\bibfnamefont
  {G.}~\bibnamefont {Rauhut}}, \bibinfo {author} {\bibfnamefont {K.~R.}\
  \bibnamefont {Shamasundar}}, \bibinfo {author} {\bibfnamefont {T.~B.}\
  \bibnamefont {Adler}}, \bibinfo {author} {\bibfnamefont {R.~D.}\ \bibnamefont
  {Amos}}, \bibinfo {author} {\bibfnamefont {A.}~\bibnamefont {Bernhardsson}},
  \bibinfo {author} {\bibfnamefont {A.}~\bibnamefont {Berning}}, \bibinfo
  {author} {\bibfnamefont {D.~L.}\ \bibnamefont {Cooper}}, \bibinfo {author}
  {\bibfnamefont {M.~J.~O.}\ \bibnamefont {Deegan}}, \bibinfo {author}
  {\bibfnamefont {A.~J.}\ \bibnamefont {Dobbyn}}, \bibinfo {author}
  {\bibfnamefont {F.}~\bibnamefont {Eckert}}, \bibinfo {author} {\bibfnamefont
  {E.}~\bibnamefont {Goll}}, \bibinfo {author} {\bibfnamefont {C.}~\bibnamefont
  {Hampel}}, \bibinfo {author} {\bibfnamefont {A.}~\bibnamefont {Hesselmann}},
  \bibinfo {author} {\bibfnamefont {G.}~\bibnamefont {Hetzer}}, \bibinfo
  {author} {\bibfnamefont {T.}~\bibnamefont {Hrenar}}, \bibinfo {author}
  {\bibfnamefont {G.}~\bibnamefont {Jansen}}, \bibinfo {author} {\bibfnamefont
  {C.}~\bibnamefont {K\"oppl}}, \bibinfo {author} {\bibfnamefont
  {Y.}~\bibnamefont {Liu}}, \bibinfo {author} {\bibfnamefont {A.~W.}\
  \bibnamefont {Lloyd}}, \bibinfo {author} {\bibfnamefont {R.~A.}\ \bibnamefont
  {Mata}}, \bibinfo {author} {\bibfnamefont {A.~J.}\ \bibnamefont {May}},
  \bibinfo {author} {\bibfnamefont {S.~J.}\ \bibnamefont {McNicholas}},
  \bibinfo {author} {\bibfnamefont {W.}~\bibnamefont {Meyer}}, \bibinfo
  {author} {\bibfnamefont {M.~E.}\ \bibnamefont {Mura}}, \bibinfo {author}
  {\bibfnamefont {A.}~\bibnamefont {Nicklass}}, \bibinfo {author}
  {\bibfnamefont {D.~P.}\ \bibnamefont {O'Neill}}, \bibinfo {author}
  {\bibfnamefont {P.}~\bibnamefont {Palmieri}}, \bibinfo {author}
  {\bibfnamefont {D.}~\bibnamefont {Peng}}, \bibinfo {author} {\bibfnamefont
  {K.}~\bibnamefont {Pfl\"uger}}, \bibinfo {author} {\bibfnamefont
  {R.}~\bibnamefont {Pitzer}}, \bibinfo {author} {\bibfnamefont
  {M.}~\bibnamefont {Reiher}}, \bibinfo {author} {\bibfnamefont
  {T.}~\bibnamefont {Shiozaki}}, \bibinfo {author} {\bibfnamefont
  {H.}~\bibnamefont {Stoll}}, \bibinfo {author} {\bibfnamefont {A.~J.}\
  \bibnamefont {Stone}}, \bibinfo {author} {\bibfnamefont {R.}~\bibnamefont
  {Tarroni}}, \bibinfo {author} {\bibfnamefont {T.}~\bibnamefont
  {Thorsteinsson}}, \ and\ \bibinfo {author} {\bibfnamefont {M.}~\bibnamefont
  {Wang}},\ }\href@noop {} {\enquote {\bibinfo {title} {{MOLPRO}, version
  2015.1, a package of ab initio programs},}\ } (\bibinfo {year} {2015}),\
  \bibinfo {note} {see http://www.molpro.net}\BibitemShut {NoStop}%
\bibitem [{\citenamefont {Spielfiedel}(2003)}]{SPIELFIEDEL2003162}%
  \BibitemOpen
  \bibfield  {author} {\bibinfo {author} {\bibfnamefont {A.}~\bibnamefont
  {Spielfiedel}},\ }\href {\doibase
  https://doi.org/10.1016/S0022-2852(02)00043-7} {\bibfield  {journal}
  {\bibinfo  {journal} {Journal of Molecular Spectroscopy}\ }\textbf {\bibinfo
  {volume} {217}},\ \bibinfo {pages} {162} (\bibinfo {year}
  {2003})}\BibitemShut {NoStop}%
\bibitem [{\citenamefont {Dunning}(1989)}]{Dunning_1989}%
  \BibitemOpen
  \bibfield  {author} {\bibinfo {author} {\bibfnamefont {T.~H.}\ \bibnamefont
  {Dunning}},\ }\href {\doibase 10.1063/1.456153} {\bibfield  {journal}
  {\bibinfo  {journal} {The Journal of Chemical Physics}\ }\textbf {\bibinfo
  {volume} {90}},\ \bibinfo {pages} {1007} (\bibinfo {year}
  {1989})}\BibitemShut {NoStop}%
\bibitem [{\citenamefont {Kendall}\ \emph {et~al.}(1992)\citenamefont
  {Kendall}, \citenamefont {Dunning},\ and\ \citenamefont
  {Harrison}}]{Dunning_1992}%
  \BibitemOpen
  \bibfield  {author} {\bibinfo {author} {\bibfnamefont {R.~A.}\ \bibnamefont
  {Kendall}}, \bibinfo {author} {\bibfnamefont {T.~H.}\ \bibnamefont
  {Dunning}}, \ and\ \bibinfo {author} {\bibfnamefont {R.~J.}\ \bibnamefont
  {Harrison}},\ }\href {\doibase 10.1063/1.462569} {\bibfield  {journal}
  {\bibinfo  {journal} {The Journal of Chemical Physics}\ }\textbf {\bibinfo
  {volume} {96}},\ \bibinfo {pages} {6796} (\bibinfo {year}
  {1992})}\BibitemShut {NoStop}%
\bibitem [{\citenamefont {Kaufmann}\ \emph {et~al.}(1989)\citenamefont
  {Kaufmann}, \citenamefont {Baumeister},\ and\ \citenamefont
  {Jungen}}]{Kaufmann}%
  \BibitemOpen
  \bibfield  {author} {\bibinfo {author} {\bibfnamefont {K.}~\bibnamefont
  {Kaufmann}}, \bibinfo {author} {\bibfnamefont {W.}~\bibnamefont
  {Baumeister}}, \ and\ \bibinfo {author} {\bibfnamefont {M.}~\bibnamefont
  {Jungen}},\ }\href {\doibase 10.1088/0953-4075/22/14/007} {\bibfield
  {journal} {\bibinfo  {journal} {Journal of Physics B: Atomic, Molecular and
  Optical Physics}\ }\textbf {\bibinfo {volume} {22}},\ \bibinfo {pages} {2223}
  (\bibinfo {year} {1989})}\BibitemShut {NoStop}%
\bibitem [{\citenamefont {Bello}\ \emph {et~al.}(2021)\citenamefont {Bello},
  \citenamefont {Lucchese}, \citenamefont {Rescigno},\ and\ \citenamefont
  {McCurdy}}]{Bello_2021}%
  \BibitemOpen
  \bibfield  {author} {\bibinfo {author} {\bibfnamefont {R.~Y.}\ \bibnamefont
  {Bello}}, \bibinfo {author} {\bibfnamefont {R.~R.}\ \bibnamefont {Lucchese}},
  \bibinfo {author} {\bibfnamefont {T.~N.}\ \bibnamefont {Rescigno}}, \ and\
  \bibinfo {author} {\bibfnamefont {C.~W.}\ \bibnamefont {McCurdy}},\ }\href
  {\doibase 10.1103/PhysRevResearch.3.013228} {\bibfield  {journal} {\bibinfo
  {journal} {Phys. Rev. Research}\ }\textbf {\bibinfo {volume} {3}},\ \bibinfo
  {pages} {013228} (\bibinfo {year} {2021})}\BibitemShut {NoStop}%
\bibitem [{\citenamefont {Kelkensberg}\ \emph {et~al.}(2009)\citenamefont
  {Kelkensberg}, \citenamefont {Lefebvre}, \citenamefont {Siu}, \citenamefont
  {Ghafur}, \citenamefont {{Nguyen-Dang}}, \citenamefont {Atabek},
  \citenamefont {Keller}, \citenamefont {Serov}, \citenamefont {Johnsson},
  \citenamefont {Swoboda}, \citenamefont {Remetter}, \citenamefont
  {L'Huillier}, \citenamefont {Zherebtsov}, \citenamefont {Sansone},
  \citenamefont {Benedetti}, \citenamefont {Ferrari}, \citenamefont {Nisoli},
  \citenamefont {L{\'e}pine}, \citenamefont {Kling},\ and\ \citenamefont
  {Vrakking}}]{kelkensberg_molecular_2009}%
  \BibitemOpen
  \bibfield  {author} {\bibinfo {author} {\bibfnamefont {F.}~\bibnamefont
  {Kelkensberg}}, \bibinfo {author} {\bibfnamefont {C.}~\bibnamefont
  {Lefebvre}}, \bibinfo {author} {\bibfnamefont {W.}~\bibnamefont {Siu}},
  \bibinfo {author} {\bibfnamefont {O.}~\bibnamefont {Ghafur}}, \bibinfo
  {author} {\bibfnamefont {T.~T.}\ \bibnamefont {{Nguyen-Dang}}}, \bibinfo
  {author} {\bibfnamefont {O.}~\bibnamefont {Atabek}}, \bibinfo {author}
  {\bibfnamefont {A.}~\bibnamefont {Keller}}, \bibinfo {author} {\bibfnamefont
  {V.}~\bibnamefont {Serov}}, \bibinfo {author} {\bibfnamefont
  {P.}~\bibnamefont {Johnsson}}, \bibinfo {author} {\bibfnamefont
  {M.}~\bibnamefont {Swoboda}}, \bibinfo {author} {\bibfnamefont
  {T.}~\bibnamefont {Remetter}}, \bibinfo {author} {\bibfnamefont
  {A.}~\bibnamefont {L'Huillier}}, \bibinfo {author} {\bibfnamefont
  {S.}~\bibnamefont {Zherebtsov}}, \bibinfo {author} {\bibfnamefont
  {G.}~\bibnamefont {Sansone}}, \bibinfo {author} {\bibfnamefont
  {E.}~\bibnamefont {Benedetti}}, \bibinfo {author} {\bibfnamefont
  {F.}~\bibnamefont {Ferrari}}, \bibinfo {author} {\bibfnamefont
  {M.}~\bibnamefont {Nisoli}}, \bibinfo {author} {\bibfnamefont
  {F.}~\bibnamefont {L{\'e}pine}}, \bibinfo {author} {\bibfnamefont {M.~F.}\
  \bibnamefont {Kling}}, \ and\ \bibinfo {author} {\bibfnamefont {M.~J.~J.}\
  \bibnamefont {Vrakking}},\ }\href@noop {} {\bibfield  {journal} {\bibinfo
  {journal} {Phys. Rev. Lett.}\ }\textbf {\bibinfo {volume} {103}} (\bibinfo
  {year} {2009})}\BibitemShut {NoStop}%
\bibitem [{\citenamefont {Furukawa}\ \emph {et~al.}(2010)\citenamefont
  {Furukawa}, \citenamefont {Nabekawa}, \citenamefont {Okino}, \citenamefont
  {Saugout}, \citenamefont {Yamanouchi},\ and\ \citenamefont
  {Midorikawa}}]{furukawa_nonlinear_2010}%
  \BibitemOpen
  \bibfield  {author} {\bibinfo {author} {\bibfnamefont {Y.}~\bibnamefont
  {Furukawa}}, \bibinfo {author} {\bibfnamefont {Y.}~\bibnamefont {Nabekawa}},
  \bibinfo {author} {\bibfnamefont {T.}~\bibnamefont {Okino}}, \bibinfo
  {author} {\bibfnamefont {S.}~\bibnamefont {Saugout}}, \bibinfo {author}
  {\bibfnamefont {K.}~\bibnamefont {Yamanouchi}}, \ and\ \bibinfo {author}
  {\bibfnamefont {K.}~\bibnamefont {Midorikawa}},\ }\href@noop {} {\bibfield
  {journal} {\bibinfo  {journal} {Phys. Rev. A}\ }\textbf {\bibinfo {volume}
  {82}} (\bibinfo {year} {2010})}\BibitemShut {NoStop}%
\bibitem [{\citenamefont {Posthumus}(2004)}]{posthumus_dynamics_2004}%
  \BibitemOpen
  \bibfield  {author} {\bibinfo {author} {\bibfnamefont {J.~H.}\ \bibnamefont
  {Posthumus}},\ }\href@noop {} {\bibfield  {journal} {\bibinfo  {journal}
  {Rep. Prog. Phys.}\ }\textbf {\bibinfo {volume} {67}},\ \bibinfo {pages}
  {623} (\bibinfo {year} {2004})}\BibitemShut {NoStop}%
\bibitem [{\citenamefont {Cattaneo}\ \emph {et~al.}(2018)\citenamefont
  {Cattaneo}, \citenamefont {Vos}, \citenamefont {Bello}, \citenamefont
  {Palacios}, \citenamefont {Heuser}, \citenamefont {Pedrelli}, \citenamefont
  {Lucchini}, \citenamefont {Cirelli}, \citenamefont {Mart{\'i}n},\ and\
  \citenamefont {Keller}}]{cattaneo_attosecond_2018}%
  \BibitemOpen
  \bibfield  {author} {\bibinfo {author} {\bibfnamefont {L.}~\bibnamefont
  {Cattaneo}}, \bibinfo {author} {\bibfnamefont {J.}~\bibnamefont {Vos}},
  \bibinfo {author} {\bibfnamefont {R.~Y.}\ \bibnamefont {Bello}}, \bibinfo
  {author} {\bibfnamefont {A.}~\bibnamefont {Palacios}}, \bibinfo {author}
  {\bibfnamefont {S.}~\bibnamefont {Heuser}}, \bibinfo {author} {\bibfnamefont
  {L.}~\bibnamefont {Pedrelli}}, \bibinfo {author} {\bibfnamefont
  {M.}~\bibnamefont {Lucchini}}, \bibinfo {author} {\bibfnamefont
  {C.}~\bibnamefont {Cirelli}}, \bibinfo {author} {\bibfnamefont
  {F.}~\bibnamefont {Mart{\'i}n}}, \ and\ \bibinfo {author} {\bibfnamefont
  {U.}~\bibnamefont {Keller}},\ }\href@noop {} {\bibfield  {journal} {\bibinfo
  {journal} {Nat. Phys.}\ }\textbf {\bibinfo {volume} {14}},\ \bibinfo {pages}
  {733} (\bibinfo {year} {2018})}\BibitemShut {NoStop}%
\bibitem [{\citenamefont {Balakrishnan}\ \emph {et~al.}(1994)\citenamefont
  {Balakrishnan}, \citenamefont {Smith},\ and\ \citenamefont
  {Stoicheff}}]{balakrishnan_dissociation_1994}%
  \BibitemOpen
  \bibfield  {author} {\bibinfo {author} {\bibfnamefont {A.}~\bibnamefont
  {Balakrishnan}}, \bibinfo {author} {\bibfnamefont {V.}~\bibnamefont {Smith}},
  \ and\ \bibinfo {author} {\bibfnamefont {B.~P.}\ \bibnamefont {Stoicheff}},\
  }\href@noop {} {\bibfield  {journal} {\bibinfo  {journal} {Phys. Rev. A}\
  }\textbf {\bibinfo {volume} {49}},\ \bibinfo {pages} {2460} (\bibinfo {year}
  {1994})}\BibitemShut {NoStop}%
\bibitem [{\citenamefont {Hogle}\ \emph {et~al.}(2015)\citenamefont {Hogle},
  \citenamefont {Tong}, \citenamefont {Martin}, \citenamefont {Murnane},
  \citenamefont {Kapteyn},\ and\ \citenamefont
  {Ranitovic}}]{hogle_attosecond_2015}%
  \BibitemOpen
  \bibfield  {author} {\bibinfo {author} {\bibfnamefont {C.~W.}\ \bibnamefont
  {Hogle}}, \bibinfo {author} {\bibfnamefont {X.~M.}\ \bibnamefont {Tong}},
  \bibinfo {author} {\bibfnamefont {L.}~\bibnamefont {Martin}}, \bibinfo
  {author} {\bibfnamefont {M.~M.}\ \bibnamefont {Murnane}}, \bibinfo {author}
  {\bibfnamefont {H.~C.}\ \bibnamefont {Kapteyn}}, \ and\ \bibinfo {author}
  {\bibfnamefont {P.}~\bibnamefont {Ranitovic}},\ }\href@noop {} {\bibfield
  {journal} {\bibinfo  {journal} {Phys. Rev. Lett.}\ }\textbf {\bibinfo
  {volume} {115}} (\bibinfo {year} {2015})}\BibitemShut {NoStop}%
\bibitem [{\citenamefont {Serov}\ and\ \citenamefont
  {Kheifets}(2014)}]{serov_p_2014}%
  \BibitemOpen
  \bibfield  {author} {\bibinfo {author} {\bibfnamefont {V.~V.}\ \bibnamefont
  {Serov}}\ and\ \bibinfo {author} {\bibfnamefont {A.~S.}\ \bibnamefont
  {Kheifets}},\ }\href@noop {} {\bibfield  {journal} {\bibinfo  {journal}
  {Phys. Rev. A}\ }\textbf {\bibinfo {volume} {89}} (\bibinfo {year}
  {2014})}\BibitemShut {NoStop}%
\bibitem [{\citenamefont {Waitz}\ \emph {et~al.}(2016)\citenamefont {Waitz},
  \citenamefont {Aslit{\"u}rk}, \citenamefont {Wechselberger}, \citenamefont
  {Gill}, \citenamefont {Rist}, \citenamefont {Wiegandt}, \citenamefont
  {Goihl}, \citenamefont {Kastirke}, \citenamefont {Weller}, \citenamefont
  {Bauer}, \citenamefont {Metz}, \citenamefont {Sturm}, \citenamefont
  {Voigtsberger}, \citenamefont {Zeller}, \citenamefont {Trinter},
  \citenamefont {Schiwietz}, \citenamefont {Weber}, \citenamefont {Williams},
  \citenamefont {Sch{\"o}ffler}, \citenamefont {Schmidt}, \citenamefont
  {Jahnke},\ and\ \citenamefont {D{\"o}rner}}]{waitz_electron_2016}%
  \BibitemOpen
  \bibfield  {author} {\bibinfo {author} {\bibfnamefont {M.}~\bibnamefont
  {Waitz}}, \bibinfo {author} {\bibfnamefont {D.}~\bibnamefont {Aslit{\"u}rk}},
  \bibinfo {author} {\bibfnamefont {N.}~\bibnamefont {Wechselberger}}, \bibinfo
  {author} {\bibfnamefont {H.~K.}\ \bibnamefont {Gill}}, \bibinfo {author}
  {\bibfnamefont {J.}~\bibnamefont {Rist}}, \bibinfo {author} {\bibfnamefont
  {F.}~\bibnamefont {Wiegandt}}, \bibinfo {author} {\bibfnamefont
  {C.}~\bibnamefont {Goihl}}, \bibinfo {author} {\bibfnamefont
  {G.}~\bibnamefont {Kastirke}}, \bibinfo {author} {\bibfnamefont
  {M.}~\bibnamefont {Weller}}, \bibinfo {author} {\bibfnamefont
  {T.}~\bibnamefont {Bauer}}, \bibinfo {author} {\bibfnamefont
  {D.}~\bibnamefont {Metz}}, \bibinfo {author} {\bibfnamefont {F.~P.}\
  \bibnamefont {Sturm}}, \bibinfo {author} {\bibfnamefont {J.}~\bibnamefont
  {Voigtsberger}}, \bibinfo {author} {\bibfnamefont {S.}~\bibnamefont
  {Zeller}}, \bibinfo {author} {\bibfnamefont {F.}~\bibnamefont {Trinter}},
  \bibinfo {author} {\bibfnamefont {G.}~\bibnamefont {Schiwietz}}, \bibinfo
  {author} {\bibfnamefont {T.}~\bibnamefont {Weber}}, \bibinfo {author}
  {\bibfnamefont {J.~B.}\ \bibnamefont {Williams}}, \bibinfo {author}
  {\bibfnamefont {M.~S.}\ \bibnamefont {Sch{\"o}ffler}}, \bibinfo {author}
  {\bibfnamefont {L.~P.~H.}\ \bibnamefont {Schmidt}}, \bibinfo {author}
  {\bibfnamefont {T.}~\bibnamefont {Jahnke}}, \ and\ \bibinfo {author}
  {\bibfnamefont {R.}~\bibnamefont {D{\"o}rner}},\ }\href@noop {} {\bibfield
  {journal} {\bibinfo  {journal} {Phys. Rev. Lett.}\ }\textbf {\bibinfo
  {volume} {116}} (\bibinfo {year} {2016})}\BibitemShut {NoStop}%
\bibitem [{\citenamefont {{Heck et al}}()}]{heck_et_al_symmetry_nodate}%
  \BibitemOpen
  \bibfield  {author} {\bibinfo {author} {\bibnamefont {{Heck et al}}},\
  }\href@noop {} {\bibinfo  {journal} {Phys Rev Res - Accept. Publ.}\
  }\BibitemShut {NoStop}%
\bibitem [{\citenamefont {Efron}(1987)}]{efronBetterBootstrapConfidence1987}%
  \BibitemOpen
\bibfield  {journal} {  }\bibfield  {author} {\bibinfo {author} {\bibfnamefont
  {B.}~\bibnamefont {Efron}},\ }\href {\doibase 10.1080/01621459.1987.10478410}
  {\bibfield  {journal} {\bibinfo  {journal} {Journal of the American
  Statistical Association}\ }\textbf {\bibinfo {volume} {82}},\ \bibinfo
  {pages} {171} (\bibinfo {year} {1987})}\BibitemShut {NoStop}%
\end{thebibliography}%

\end{document}